\documentclass[english,reprint, aps,prx,floatfix]{revtex4-2}
\usepackage[T1]{fontenc}
\usepackage[latin9]{inputenc}
\usepackage{float}
\usepackage{verbatim}
\usepackage{float}
\usepackage{amsmath}
\usepackage{amssymb}
\usepackage{graphicx}
\usepackage{color}
\usepackage[dvipsnames]{xcolor}
\usepackage[colorlinks=true,linkcolor=blue]{hyperref}
\hypersetup{
    colorlinks = true,
    linkbordercolor = {white},
    linkcolor= BrickRed, 
    citecolor=blue,	 
    urlcolor=blue,
    }

\makeatletter

\renewcommand\[{\begin{equation}}
\renewcommand\]{\end{equation}} 

\makeatother

\usepackage{babel}
\begin{document}
\title{Exact description of quantum stochastic models as quantum resistors}

\author{Tony Jin$^1$, Jo\~ao S. Ferreira$^1$, Michele Filippone$^{1,2}$
and Thierry Giamarchi$^1$}
\affiliation{$^1$Department of Quantum Matter Physics, Ecole de Physique
University of Geneva, Quai Ernest-Ansermet 24, CH-1211 Geneva 4, Switzerland}
\affiliation{$^2$Universit\'e   Grenoble  Alpes,  CEA,  IRIG-MEM-L\_Sim,  F-38000,  Grenoble,  France}

\global\long\def\ket#1{\left| #1\right\rangle }%
\global\long\def\bra#1{\left\langle #1 \right|}%
\global\long\def\kket#1{\left\Vert #1\right\rangle }%
\global\long\def\bbra#1{\left\langle #1\right\Vert }%
\global\long\def\braket#1#2{\left\langle #1\right.\left|#2\right\rangle }%
\global\long\def\bbrakket#1#2{\left\langle #1\right.\left\Vert #2\right\rangle }%
\global\long\def\av#1{\left\langle #1\right\rangle }%
\global\long\def\Tr{\text{Tr}}%
\global\long\def\id{\mathbb{I}}%
\global\long\def\inf{\infty}%
\global\long\def\a{\alpha}%
\global\long\def\b{\beta}%
\global\long\def\g{\gamma}%
\global\long\def\G{\Gamma}%
\global\long\def\w{\omega}%
\global\long\def\t{\tau}%
\global\long\def\O{\Omega}%
\global\long\def\d{\delta}%
\global\long\def\D{\Delta}%
\global\long\def\m{\mu}%
\global\long\def\LL{\mathcal{L}}%
\global\long\def\H{\mathcal{H}}%
\global\long\def\S{\Sigma}%
\global\long\def\s{\sigma}%
\global\long\def\r{\rho}%
\global\long\def\l{\lambda}%
\global\long\def\dag{\dagger}%
\global\long\def\e{\epsilon}%
\global\long\def\x{\chi}%
\global\long\def\p{\phi}%
\global\long\def\th{\theta}%
\global\long\def\ra{\rightarrow}%
\global\long\def\fp#1#2#3#4#5#6{\left(\psi_{#1}^{#3}\bar{\psi}_{#1}^{#4}\psi_{#1+1}^{#5}\bar{\psi}_{#1+1}^{#6}\right)_{#2}}%
\global\long\def\psib{\bar{\psi}}%

\begin{abstract} 
We study the transport properties of generic out-of-equilibrium quantum systems connected to fermionic reservoirs. We develop a new perturbation scheme in the inverse system size, named $1/N$ expansion, to study a large class of out of equilibrium diffusive/ohmic systems. The bare theory is described by a Gaussian action corresponding to a set of independent two level systems at equilibrium. This allows a simple and compact derivation of the diffusive current as a first order pertubative term. In addition, we obtain exact solutions for a large class of quantum stochastic Hamiltonians (QSHs) with time and space dependent noise, using a self consistent Born diagrammatic method in the Keldysh representation. We show that these QSHs exhibit diffusive regimes which are encoded in the Keldysh component of the single particle Green's function. The exact solution for these QSHs models confirms the validity of our system size expansion ansatz, and its efficiency in capturing the transport properties. We consider in particular three fermionic models: {\it i)} a model with local dephasing {\it ii)} the quantum simple symmetric exclusion process model {\it iii)} a model with long-range stochastic hopping. For {\it i)} and {\it ii)} we compute the full temperature and dephasing dependence of the conductance of the system, both for two- and four-points measurements. Our solution gives access to the regime of finite temperature of the reservoirs 
which could not be obtained by previous approaches. For {\it iii)}, we unveil a novel ballistic-to-diffusive transition governed by the range and the nature (quantum or classical) of the hopping. As a by-product, our approach equally describes the mean behavior of quantum systems under continuous measurement.
\end{abstract}

\maketitle

\section{\label{sec:intro}Introduction}

Diffusion is the transport phenomenon  most commonly encountered in nature. It implies that globally conserved quantities such as energy, charge, spin or mass spread uniformly all over the system according to Fick/Ohm's law
\begin{equation}\label{eq:fick}
 J=- D\nabla n\,,
\end{equation}
where  the diffusion constant $D$ relates the current density $J$ to a superimposed density gradient $\nabla n$. 

Despite its ubiquity, understanding the emergence of classical diffusive phenomena from underlying quantum mechanical principles is highly non trivial. Early works based on field theory and perturbative methods \cite{giamarchi_umklapp_1d,rosch_conservation_1d} pointed out the possibility that interactions do not necessarily lead to diffusion at finite temperature, a question addressed then more rigorously by using the concepts of integrability \cite{zotos_conductivity_integrable}. These questions have then fueled many exciting discoveries in low-dimensional interacting systems~\cite{bertinifinite2021}. 
A notable example is the ballistic-to-diffusive transition  in quantum integrable XXZ spin chains~\cite{zotosFiniteTemperatureDrude1999,Prosen2011,Ljubotina2017,Ilievski2018,DeNardis2020,DeNardis2018}, which also exhibit a superdiffusive point in the Kardar-Parisi-Zhang universality class~\cite{kardarDynamicScalingGrowing1986,kriecherbauerPedestrianTextquotesinglesView2010,ljubotina2019kardar,Gopalakrishnan2019,denardisSuperdiffusionEmergentClassical2020}.
These discoveries have motivated the generalized hydrodynamical descriptions of integrable systems~\cite{castro-alvaredoEmergentHydrodynamicsIntegrable2016,bertiniTransportOutofEquilibriumXXZ2016}, providing an elegant path to the question of diffusion at finite temperature~\cite{Ilievski2017}, and paving the way to the description of  diffusive phenomena based on perturbative approaches~\cite{denardisDiffusionGeneralizedHydrodynamics2019,friedmanDiffusiveHydrodynamicsIntegrability2020,znidaricNonequilibriumSteadystateKubo2019a,znidaricWeakIntegrabilityBreaking2020a,znidaricAbsenceSuperdiffusionQuasiperiodic2020,ferreiraBallistictodiffusiveTransitionSpin2020a}. 

\begin{figure}
\begin{centering}
\includegraphics[width=\columnwidth]{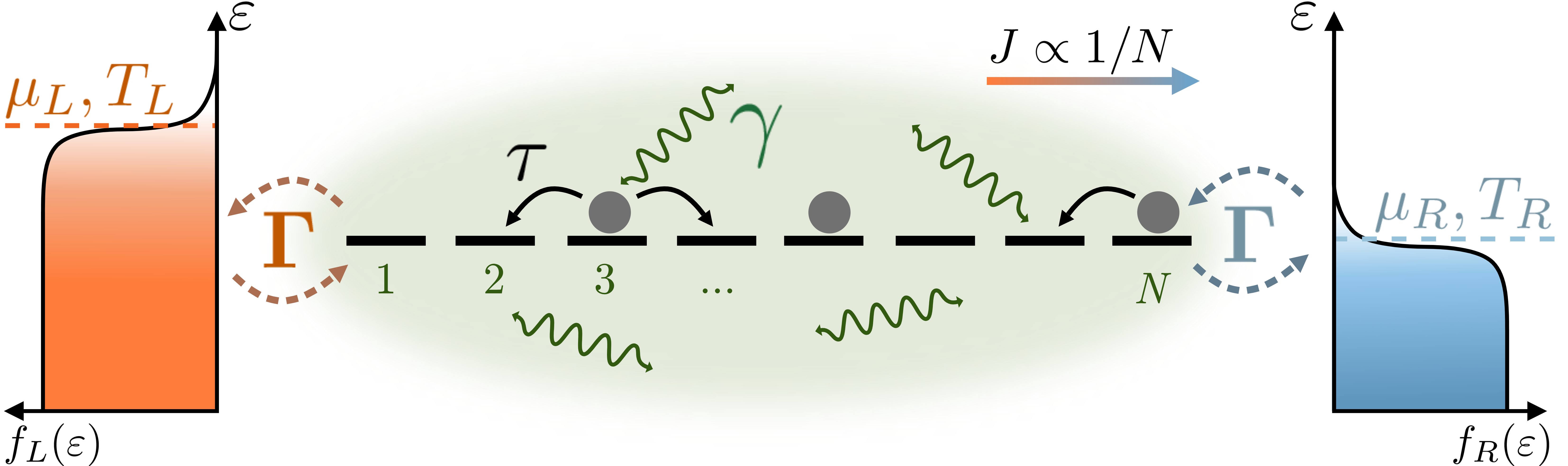}
\end{centering}
\caption{\label{fig:Generic-setup}
A stationary current $J$ flows in a  one-dimensional lattice when connected to left (L) and right (R) fermionic reservoirs, described by Fermi distributions $f(\varepsilon)$ with different temperatures $T$ or chemical potentials $\mu$. The wiggly lines denote dissipative degrees of freedom acting on the system with rate $\gamma$. For a fixed difference of chemical potential $\delta\mu=\mu_L-\mu_R$, dissipative terms are normally responsible for the Ohmic suppression of the current, $J\propto 1/N$. }.  
\end{figure}

The out-of-equilibrium driving protocol illustrated in Fig.~\ref{fig:Generic-setup}, where a system is coupled to external dissipative baths, has been crucial to unveil and characterize such exotic transport phenomena~\cite{Prosen2011,znidaricSpinTransportOneDimensional2011,Ljubotina2017,yamanaka2021exact}. 
It allows to study disordered systems~\cite{znidaricDiffusiveSubdiffusiveSpin2016,mendoza-arenasAsymmetryEnergySpin2019,znidaricInteractionInstabilityLocalization2018}, uncover novel integrable  structures~\citep{Prosen2011,ProsenEssler_Mapping}, and show diffusive transport
\citep{Znidaric__XXdeph,Znidaric__dephasingXXZ,Eisler_CrossoverBallisticDiffusive,BauerBernardJin_Stoqdissipative,BernardJin_QSSEP,BastianelloDeNardisDeluca_GHDDeph}. These open quantum systems~\citep{GardinerZoller_quantumnoise,BreuerPetruccione_book,Breuer_HeattransportCPmaps}, are described within the Lindblad formalism~\cite{Gorini1976,Lindblad1976}, which is actively employed to investigate the exotic dynamics induced by non-trivial  interactions with external degrees of freedom, such as lattice vibrations, quantum
measurements~\cite{skinnerMeasurementInducedPhaseTransitions2019,albertonTrajectoryDependentEntanglement2020,buchholdEffectiveTheoryMeasurementInduced2021,caoEntanglementFermionChain2019,muller2021measurementinduced,zhang2021universal}, dephasing~\cite{tonielliOrthogonalityCatastropheDissipative2019a,mitchisonSituThermometryCold2020a,dolgirevNonGaussianCorrelationsImprinted2020a,albaUnboundedEntanglementProduction2021,wolffEvolutionTwotimeCorrelations2019,lacerda2021dephasing}, losses~\cite{robertsDrivenDissipativeQuantumKerr2020,fromlFluctuationInducedQuantumZeno2019a,rossini2020strong,rosso2021one,yamamoto_2019,muller2021shape}, coupling to a lightfield~\cite{dograDissipationinducedStructuralInstability2019,pichlerNonequilibriumDynamicsBosonic2010,halatiNumericallyExactTreatment2020}
and environmental engineering~\cite{verstraete2009quantum}. 
 
This research activity is also motivating ongoing experiments, where  recent progress in space- and time-resolved techniques is applied to  directly observe  emergent  diffusive and exotic dynamics  in various quantum systems, including cold atoms~\cite{sommerUniversalSpinTransport2011,dograDissipationinducedStructuralInstability2019,jepsenSpinTransportTunable2020a,jepsenTransverseSpinDynamics2021,bouganne2020anomalous},  spin chains~\cite{takigawa1996,Thurber2001,pratt2006,maeter2013,scheieDetectionKardarParisi2021,zuEmergentHydrodynamicsStrongly2021} and solid-state~\cite{mollEvidenceHydrodynamicElectron2016,ellaSimultaneousVoltageCurrent2019,sulpizioVisualizingPoiseuilleFlow2019}. In this context, theoretical predictions are usually made case-by-case, with strong constraints on geometries and driving protocols~\cite{prosen2021}. Thus, devising versatile tools to solve generic quantum models that show diffusion becomes crucial to understand emerging classical Ohmic transport.  

In this paper, we develop a novel approach to characterize the bulk transport properties of quantum resistors which we show to be {\it exact} and {\it systematic} for a wide class of quantum stochastic Hamiltonians (QSHs). Our starting point is the Meir-Wingreen's formula~\citep{MeirWingreenformula,JinFilipponeGiamarchi_GenericMarkovian} (MW), which expresses the current  $J$ of a system driven at its boundaries, see Fig.~\ref{fig:Generic-setup}, in terms of single-particle Green's functions. We show that, for Ohmic systems, the MW formula supports an expansion of the current in terms of the inverse of the system size $N$. We illustrate how to perform practically this $1/N$ expansion, which reveals efficient to derive the diffusive current and the diffusion constant: we assume that, in the $N\rightarrow\infty$ limit, diffusive lattices admit a simple description in terms of independently equilibrated sites and demonstrate that a well-chosen perturbation theory over this trivial state leads to the desired  $1/N$ expansion. 

 We provide a comprehensive demonstration of the validity of our approach in the context of QSHs. Relying on diagrammatic methods and out-of-equilibrium field theory~\cite{Kamenev2011}, we show that single-particle Green's functions of QSHs can be exactly and systematically derived relying on the self-consistent Born approximation (SCBA) -- a generalization of previous results derived for a dephasing impurity in a thermal bath~\cite{dolgirevNonGaussianCorrelationsImprinted2020a}. Equipped with this exact solution, and relying on MW formula, we explicitly derive the dissipative current flowing in the system and show that the Keldysh component of the single particle Green's function encodes the Ohmic suppression of the current. Then, we explicitly derive the asymptotically equilibrated state  by ``coarse-graining'' of single particle Green's functions and validate our procedure to perform the $1/N$ expansion.  

We illustrate the effectiveness and versatility of our approach for three different QSHs of current interest:  {\it i}) the dephasing model~\citep{Znidaric__XXdeph,Znidaric_dephasing,Znidaric__dephasingXXZ,ProsenEssler_Mapping,Monthus_2017}; {\it ii}) the quantum symmetric simple exclusion process (QSSEP) \citep{BauerBernardJin_EquilibriumQSSEP,BernardJin_QSSEP,BernardJin_SolutionQSSEPcontinue,BernardLeDoussal_StochasticCFT,EsslerPiroli_Operatorfragmentation,BernardPiroli_QSSEPentanglement} and  {\it iii}) models with stochastic  long range hopping~\citep{Nahum_AlltoAllRandomunitary,muller2021measurementinduced}. The case studies ({\it i}) and ({\it ii})  illustrate the effectiveness of our approach, providing simple derivations of the current $J$ and  of the diffusion constant $D$, in alternative to approaches relying on matrix-product state \citep{Znidaric__XXdeph,Znidaric_dephasing,Znidaric__dephasingXXZ}, integrability~\citep{ProsenEssler_Mapping} or other case-by-case solutions \citep{BernardJin_QSSEP,Eisler_CrossoverBallisticDiffusive}. Additionally, we address previously unexplored regimes, by exactly solving the out-of-equilibrium problem with fermion reservoirs at arbitrary temperatures and chemical potentials. Our approach also allows to access two-times correlators in the stationary state which were not described by previous studies. For case ({\it iii}), we  show instead the ability of our approach to predict  novel and non-trivial transport phenomena, namely a displacement of the ballistic-to-diffusive transition induced by coherent nearest-neighbor tunneling in one-dimensional chains. A by-product of our analysis is that all the results presented here apply also for system under continuous measurement, which are currently  attracting a lot of interest in the context of measurement induced phase transition \citep{skinnerMeasurementInducedPhaseTransitions2019,Nahum_AlltoAllRandomunitary,buchholdEffectiveTheoryMeasurementInduced2021,muller2021measurementinduced}.

Our paper is structured as follows. Section~\ref{sec:Analy-D} describes how the MW formula is a good starting point to build a systematic expansion of the current in terms of the inverse system size $N$.  Section~\ref{sec:SCBA} presents QSH and shows the exactitude of SCBA for the computation of single-particle  self-energies. Section~\ref{sec:applications} shows how our formalism allows to fully compute the transport properties of the dephasing model, the QSSEP and the long-range model. Section~\ref{sec:Conclusion} is dedicated to our conclusions and the discussion of the future research perspectives opened by our work. 

\section{Resistive scaling in finite-size boundary driven systems and perturbative approach}\label{sec:Analy-D}

In this section, we introduce generic tools aimed at studying diffusive transport in boundary-driven setups like those of Fig.~\ref{fig:Generic-setup}. For these setups, the current is given by the MW formula~\citep{MeirWingreenformula}. In the simplified (yet rather general) situation, where the reservoirs have a constant density of states and the tunnel exchange of particles does not depend on energy, the MW formula reads (we assume $e=\hbar=k_B=1$): 
\begin{multline}\label{eq:MW}
J= i\int\frac{d\omega}{2\pi}\Tr\left\{\frac{1}{2}\big(\G_{L}-\G_{R}\big)G^{\cal K}+\right.\\
~~~\left. \left[\left(f_{L}-\frac{1}{2}\right)\G_{L}-\left(f_{R}-\frac{1}{2}\right)\G_{R}\right]\big(G^{\cal R}-G^{\cal A}\big)\right\}\,,
\end{multline}
where $f_{L(R)}(\omega)=[e^{(\omega-\mu_{L(R)})/T_{L(R)}}+1]^{-1}$ are the Fermi distributions associated to the left and right
reservoir with chemical potentials $\mu_{L(R)}$ and temperatures $T_{L(R)}$. $G^{\cal R/A/K}$ are  the retarded ($\cal R$), advanced ($\cal A$) and Keldysh ($\cal K$) components of the single-particle Green's functions of the system. They are defined in time representation as $G^{\cal R}_{j,k}(t-t')=-i\theta(t-t')\langle\{c_j(t),c^\dagger_k(t')\}\rangle$, $G^{\cal A}_{j,k}(t-t')=[G^{\cal R}_{j,k}(t'-t)]^*$ and $G^{\cal K}_{j,k}(t-t')=-i\langle[c_j(t),c^\dagger_k(t')]\rangle$, where the (curly)square brackets indicate (anti)commutation~\footnote{The dependence of the Green's functions on time differences $t-t'$, instead of separate times $t,t'$ is a consequence of the fact that we consider stationary situations.}. $c_j$ is the annihilation operator of a spinless fermion at site $j$. 
The $\G_{L(R)}$ matrices describe system-reservoirs couplings. 

Our aim is to establish a systematic procedure to compute diffusive current for large systems. The starting point will be the state of the system in the thermodynamic limit ($N\rightarrow\infty$). By identifying in the MW formula~\eqref{eq:MW} the terms leading to Fick's law~\eqref{eq:fick}, we motivate the simple structure of the problem for an infinite system size. In resistive systems, a fixed difference of density $\Delta n:=n_{1}-n_{N}$ at the edges of the system enforces the $1/N$ suppression of the current ($J\propto \nabla n \propto \Delta n/N$). It is thus natural to perform a perturbative $1/N$ expansion of the current on the $N\rightarrow\infty$ state. We conjecture a possible perturbation scheme and show its validity in the context of QSHs.

Without loss of generality, we focus on discrete $1D$ lattice systems of size $N$~\footnote{The extension to different geometries and additional degrees of freedom is straightforward.}. In this case, the $\Gamma_{L(R)}$ matrices in Eq.~\eqref{eq:MW} acquire a simple form in  position space: $[\G_{L(R)}]_{j,k}=\Gamma\delta_{j,1(N)}\delta_{j,k}$.   We also  express the local densities in terms of Green's functions, namely  $2n_j=2\langle c^\dagger_jc_j\rangle=1-i\int d\omega\, G_{j,j}^{\cal K}(\omega)/(2\pi)$, which also implies $2i \Delta n=G^{\cal K}_{1,1}(t=0)-G^{\cal K}_{N,N}(t=0)=\Delta G^{\cal K}$. The MW formula then acquires the more compact form:
\begin{equation} \label{eq:MWDn}
J=\Gamma \int d\omega \Big[f_L(\omega)\mathcal A_L(\omega)-f_R(\omega)\mathcal A_R(\omega)\Big]-\Gamma\Delta n\,,
\end{equation}
 where we have  introduced the local spectral densities $\mathcal A_{L(R)}(\omega)=-\frac1\pi\mbox{Im}[G^{\cal R}_{1,1(N,N)}(\omega)]$ and made use of the fact that $\int d\omega \mathcal A_{L(R)}(\omega)=1$. 
 
 The local spectral densities $\mathcal A_{L(R)}(\omega)$ exponentially converge in the thermodynamic limit $N\rightarrow\infty$. This feature is generally expected and is illustrated in Fig.~\ref{fig:spectral_dephasing} for  different classes of QSHs. 
 This observation allows to establish that the  $1/N$ scaling, proper to  diffusive currents, must entirely arise from $\Delta n$ in \eqref{eq:MWDn}. The possibility to ignore the size-dependence of the first term of \eqref{eq:MWDn} imposes strong constraints on the $1/N$ expansion of the difference of density $\Delta n$ in diffusive systems. If we write this expansion as
\begin{equation}\label{eq:GKexpansion}
2i\Delta n=\Delta G^{\cal K}=\Delta G^{(\infty)}+\frac{1}{N}\Delta G'+\ldots
\end{equation} 
one notices immediately that the leading term $\Delta G^{(\infty)}$ has to compensate the first one in \eqref{eq:MWDn}, implying 
\begin{equation}\label{eq:dG0}
\begin{split}
\frac{ \Delta G^{(\infty)}}{2i}= \int d\omega \Big[f_L(\omega)\mathcal A_{1,1}(\omega)-f_R(\omega)\mathcal A_{N,N}(\omega)\Big]\,.
\end{split}
\end{equation}
A sufficient but not necessary condition fulfilling this relation is obtained by imposing at each boundary:
\begin{equation}\label{eq:equidensities}
\int \frac{d\omega}{2\pi} G_{L(R)}^{\mathcal{K}(\infty)}(\omega)=-i\int d\omega \tanh\left(\frac{\omega-\mu_{L(R)}}{2T_{L(R)}}\right)\mathcal A_{L(R)}(\omega)\,,
\end{equation}
which will turn out to be satisfied for QSHs. These relations have a simple and interesting interpretation. In the infinite size limit, the flowing current is zero and thus the stationary value of the densities at the boundary can be computed by supposing that they fulfill a \emph{fluctuation-dissipation} relation or equivalently, that these sites are at equilibrium with the neighboring reservoirs.

Reinjecting \eqref{eq:GKexpansion} in the MW formula gives the current 
\begin{equation}\label{eq:current1}
J=i\frac{\Gamma}{2N}\Delta G'
\end{equation}
and as expected, we get the $1/N$ diffusive scaling. This relation tells us that the information about the diffusion constant is hidden in the $1/N$ correction to the density profile which can be in general a non trivial quantity to compute. However, we will see in the following that there is a shorter path to access it via the use of an infinite system size perturbation theory.

\begin{figure}
\begin{center}
\includegraphics[width=\columnwidth]{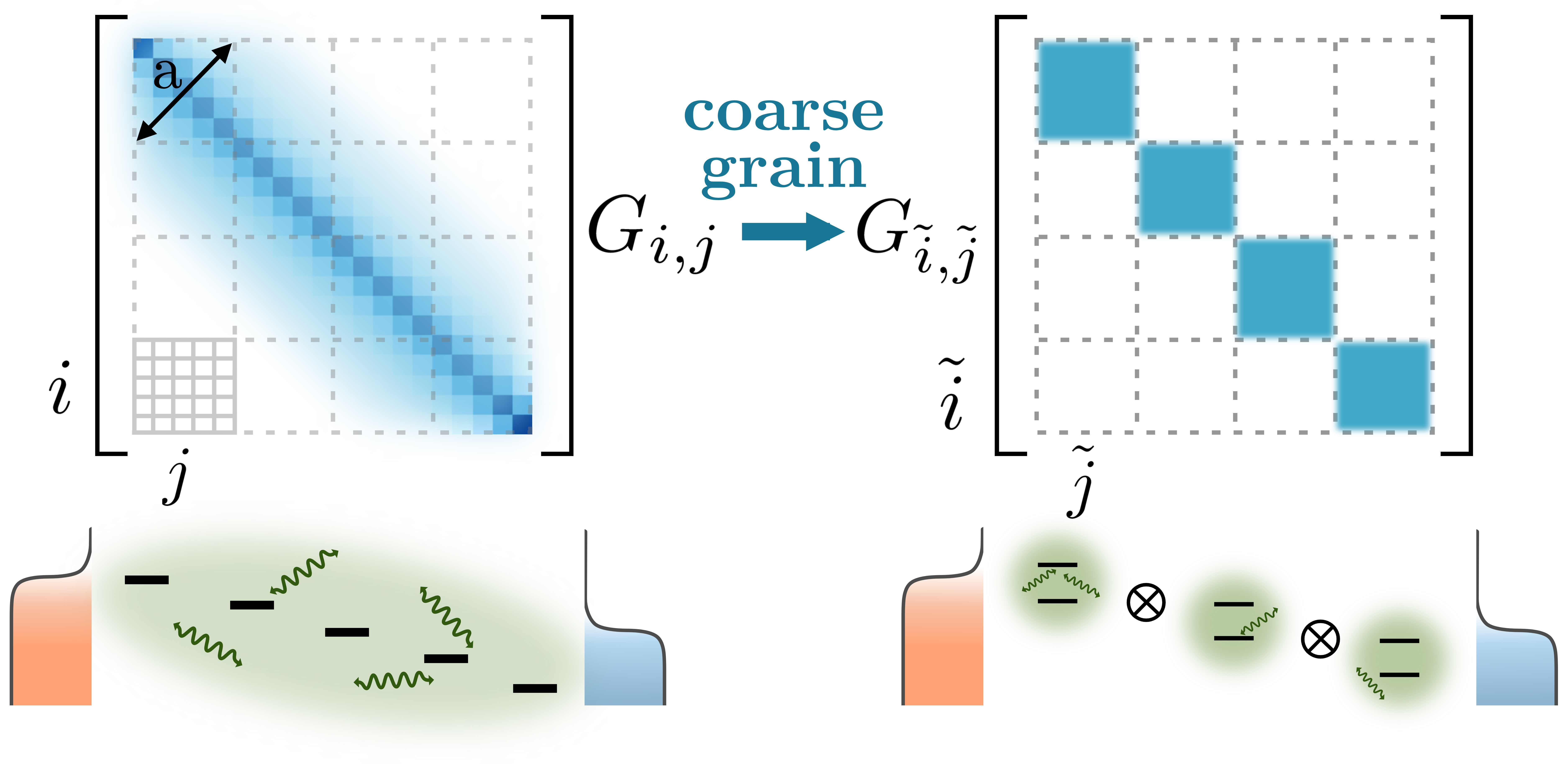}
\end{center}
\caption{\label{fig:RG_cartoon}
Cartoon picture of the coarse-graining procedure. On the left, spacial correlations in the infinite size limit are depicted. These decay exponentially as a function of the distance and are non-zero only within a finite length $a$. By coarse-graining the theory over this typical length, we obtain an effective theory (right) consisting of an ensemble of uncoupled sites with a finite self-energy at equilibrium.}
\end{figure}

The main idea of the $1/N$ perturbation is to find an effective simple theory that captures the relevant properties of the system in the $N\to\infty$ limit. From there, transport quantities are computed \emph{perturbatively} on top of this limit theory.
To determine this effective theory, we conjecture that there is a typical length $a$ beyond which two points of the systems can be considered to be statistically independent. Thus, by coarse-graining the theory over cells of size $a$, each cell becomes uncoupled and in local equilibrium, see Fig.~\ref{fig:RG_cartoon}.

The reasons motivating such factorization are twofold. First, the current is suppressed as $1/N$ in the large system size limit, so the infinite size theory should predict a null stationary current. Second, factorization of stationary correlations  has actually been demonstrated for a certain number of diffusive toy models, most notably in the context of large deviations and macroscopic fluctuation theory \citep{MFT_Review,Derrida_ReviewSSEP,BauerBernardJin_EquilibriumQSSEP,BernardJin_QSSEP}. For instance, it is known that the $n^{th}$ connected correlation functions of physical observables, such as density, generically behaves as $N^{-(n-1)}$. Thus, it is natural to assume that for $N\to \infty$, correlations must be exponentially decaying over a length $a$. We will show explicitly that in all of the examples studied, this factorization in the coarse-grained theory will turn out to be true and provide an analytic estimation for $a$ in App.\ref{sec:a-estimate}.

We now put these assumptions on formal grounds. Let $\tilde{j}$ and $\tilde{k}$ be the spatial indices of the coarse-grained theory
\begin{equation}\label{eq:coarse_graining}
G_{\tilde{j},\tilde{k}}^{\mathcal{R/A/K}}:=\frac{1}{a}\sum_{m,n=0}^{a-1}G_{\tilde{j}a+m,\tilde{k}a+n}^{\mathcal{R/A/K}}\,.
\end{equation}
The relation between the different components $\mathcal{R},\,\mathcal A$ and $\mathcal K$ of the single particle Green's functions are assumed to describe  uncoupled sites at equilibrium with a local  self-energy $\Sigma_{\tilde j}$~\cite{Kamenev2011}. These conditions require then local fluctuation-dissipation relations of the form
\begin{equation}
G_{\tilde{j},\tilde{k}}^{\mathcal{K}(\infty)}(\omega)=\delta_{\tilde{j},\tilde{k}}\tanh\left(\frac{\omega-\mu_{\tilde{j}}}{2T_{\tilde{j}}}\right)\Big[G_{\tilde{j},\tilde{j}}^{\cal R}(\omega)-G_{\tilde{j},\tilde{j}}^{\cal A}(\omega)\Big],\label{eq:GKinfinity}
\end{equation}
 with retarded and advanced Green's functions which are diagonal in the coarse-grained space representation
\begin{equation}
    G^{\cal R(A)}_{\tilde{j},\tilde{k}}(\omega)=\frac{\delta_{\tilde{j},\tilde{k}}}{\omega -\omega^0_{\tilde{j}} \pm  \Sigma_{\tilde{j}}(\omega)}\,.\label{eq:GRinfinity}
\end{equation}
These relations fix entirely the stationary property of the system in the infinite size limit. The specification of the free parameters $\mu_{\tilde{j}},T_{\tilde{j}},\omega^0_{\tilde{j}}$ and $\Sigma_{\tilde{j}}$ have to be done accordingly to the model under consideration. We will see that they take a simple form for QSHs, namely the self-energy $\Sigma_{\tilde j}$ is frequency independent and the $\mu_{\tilde j},T_{\tilde j}\gg\omega$ limit can be taken taken in Eq.~\eqref{eq:GKinfinity}, as expected in the Markovian limit of the dissipative process~\cite{JinFilipponeGiamarchi_GenericMarkovian}. 

To get the current, one needs to go one step further and understand which terms have to be expanded. The thermodynamic equilibrated theory does not exhibit transport, thus should be left invariant by the part of the Hamiltonian that commutes with the conserved quantity, for us the local particle density. It is then natural to conjecture that the perturbative term for the current is given by the dynamical part of the theory, that is, the part of the Hamiltonian $\hat{H}_{{\rm dyn}}$ which does not commute with the local  density. Thus, we conjecture that, at order $1/N$, the current is given by : 
\begin{equation} \label{eq:perturbativecurrent}
J=\langle\hat{J}\hat{H}_{\rm{dyn}}\rangle_{\infty}\,,
\end{equation}
where the $\langle\rangle_{\infty}$ means the expectation value must be taken with respect to the infinite system size theory. This formula has the remarkable advantage that its computational complexity is very low since the coarse-grained theory is Gaussian. 
We remark that the $1/N$ expansion presented here is \emph{not} a standard expansion in the hopping amplitude $\tau$, since the latter has an exponentially large degenerate manifold of states at $\tau=0$.

In Sec.~\ref{sec:applications}, we show explicitly how these ideas unfold for QSHs, by comparing computations done from the $1/N$ theory with the one obtained from the exact solution that we present in the following Section Sec.~\ref{sec:SCBA}. Understanding to which extent and under which conditions Eqs.~(\ref{eq:GKinfinity},\ref{eq:GRinfinity}) and~\eqref{eq:perturbativecurrent} can be applied is one of the very challenging direction of study, in particular in the context of interacting quantum systems without bulk dissipative terms.

\section{Validity of the self-consistent Born approximation for Quantum stochastic Hamiltonians}\label{sec:SCBA} 

In this section, we present a class of quantum stochastic models and associated Liouvillians~\eqref{eq:Liouvillian}, that describe either stochastic local dephasing or stochastic jumps of fermionic particles on a graph. The random processes are defined by a quantum Markov equation also known as a Lindblad equation. We will show explicitly two ways, exemplified by Eqs.~\eqref{eq:stochHamiltonian} and~\eqref{eq:measurement}, to associate an underlying quantum stochastic model to such Lindblad equation, a process known as \emph{unraveling} or \emph{dilatation} \citep{Dalibard_Unraveling, Carmichael_bookopensystems,Belavkin_Nondemolitionmeasurements}. Of particular interest for us is the description in terms of quantum stochastic Hamiltonians (QSHs)~\eqref{eq:stochHamiltonian}. It  provides a way to resum exactly the perturbative series associated to the stochastic noise, which coincides with the self-consistent Born approximation (SCBA) for single particle Green's functions. This method was originally devised for the particular case of a single-site dephaser in Ref.~\citep{dolgirevNonGaussianCorrelationsImprinted2020a} and we extend it here to more general situations. We will show in Section \ref{sec:applications} that, relying on SCBA, we can derive the diffusive transport properties of these models and show the validity of the assumptions underpinning the perturbative $1/N$ expansion presented in Sec.~\ref{sec:Analy-D}.

\begin{figure}
\begin{center}
\includegraphics[width=0.7\columnwidth]{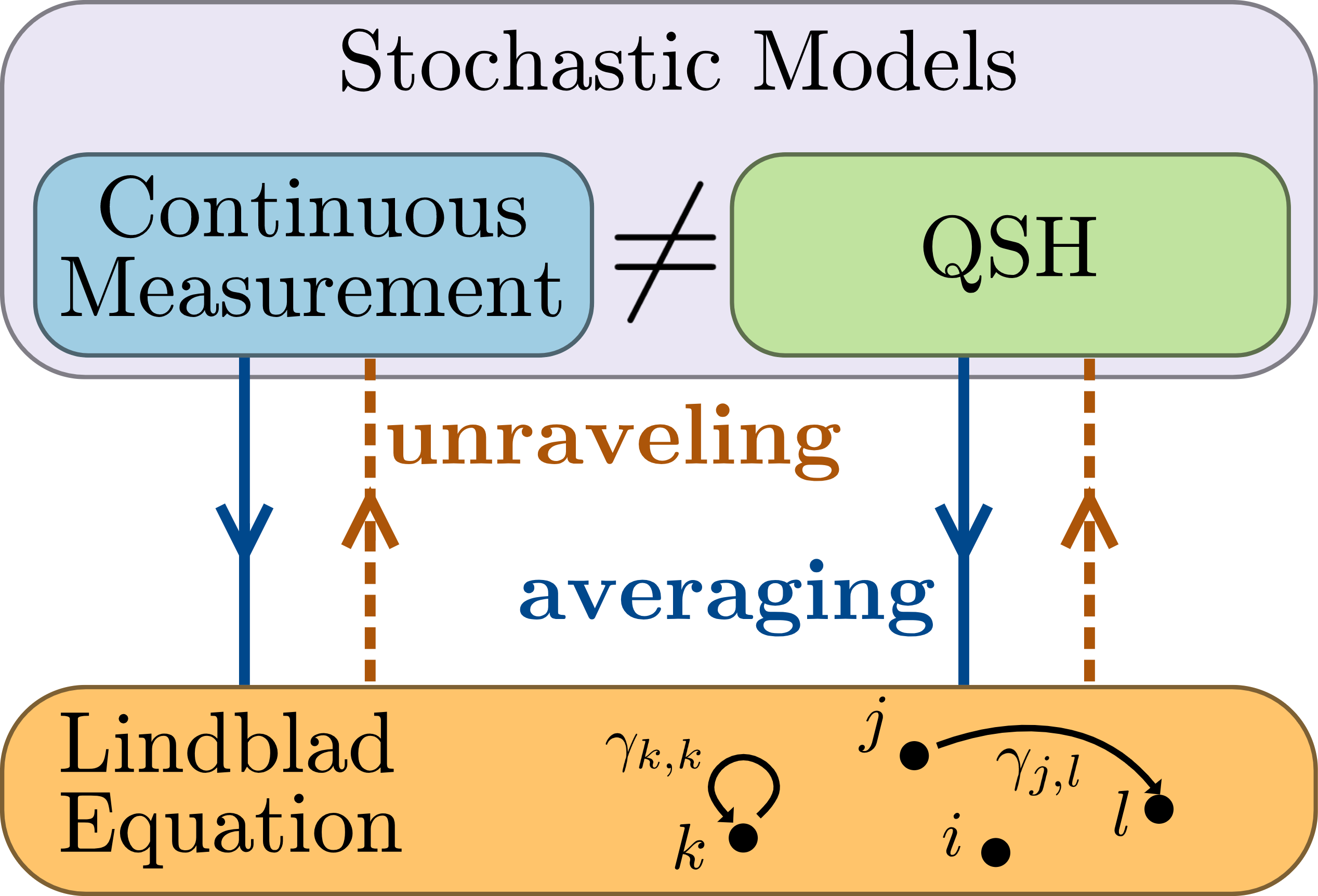}
\end{center}
\caption{\label{fig:graph}
Schematic representation of our random process. The orange box represents the Lindblad equation \eqref{eq:Liouvillian} which describes random quantum jumps between sites connected by an arrow. An arrow leaving and arriving at the same site represents a local dephasing. To a given Lindblad equation, we can associate multiple stochastic process (blue and green boxes), a process called \emph{unraveling} (orange dashed lines). The Lindblad equation is recovered by averaging over the noisy degrees of freedom (full blue lines). We show that the unraveling in terms of quantum stochastic Hamiltonian (QSH) is particularly useful for the diagrammatic expansion of the theory.}
\end{figure}

Consider a graph  made of discrete points, each corresponding to a site. To such graph we associate a Markovian process where spinless fermions on a given site can jump to any other site only if the
target site is empty, see Fig.~\ref{fig:graph}. We define $\gamma_{ij}\geq 0$ as the probability rate associated to the process of a fermion jumping from $i$ to $j$ and $\g_{ji}=\g_{ij}$ the reverse process. The generator of such process is
given by the Liouvillian, which acts on the density matrix $\rho$ of the system: 
\begin{equation}\label{eq:Liouvillian}
\begin{split}
{\cal L}(\rho)=  \sum_{i,j}&\gamma_{i,j}\left(2c_j^{\dagger}c_i\rho c_i^{\dagger}c_j-\big\{c_i^{\dagger}c_j c_j^{\dagger}c_i,\rho\big\}\right)\,.
\end{split}
\end{equation}
The total evolution of the density matrix $\rho$ is in general given by 
\begin{equation}
\frac{d}{dt}\rho={\cal L}_{0}(\rho)+{\cal L}(\rho)\,,
\end{equation}
where ${\cal L}_{0}$ generates what we call the \emph{free evolution},
in the sense that ${\cal L}_{0}$ is quadratic in the fermion operators $c_i$  and the related spectrum and propagators can be efficiently computed  with Wick's theorem~\cite{Prosen_thirdquantization,Guo2017}. Such Liouvillians can generally describe single-particle Hamiltonians or dissipative processes (coherent hopping, losses,\ldots).  We will consider
${\cal L}(\rho)$ as a perturbation on top of this theory. 

There exists a general procedure to see ${\cal L}(\rho)$ as the emergent
\emph{averaged }dynamics of an underlying  microscopic stochastic, yet Hamiltonian, 
process. Lifting ${\cal L}(\rho)$ to this stochastic process is known as  \emph{unraveling} and there is not a unique way of
doing so, see Fig.~\ref{fig:graph}. The stochastic Hamiltonian can be treated as a perturbation in field-theory which requires the summation of an infinite series. 
Our strategy is to pick the relevant stochastic theory for which there exists a simple
way to reorganize the summation and then take the average in order
to get the mean evolution. 

We now proceed to present the unraveled theory. Let $dH_{t}$ be
the stochastic Hamiltonian increment, generating  the evolution, which is defined by
\begin{equation}
\left|\psi_{t+dt}\right\rangle =e^{-idH_{t}}\left|\psi_{t}\right\rangle\,.
\end{equation}
We work in the It\={o} prescription  and consider stochastic Hamiltonians of the form 
\begin{equation}
dH_{t}=\sum_{i,j}\sqrt{2 \gamma_{i,j}}c_{j}^{\dagger} c_i dW_{t}^{i,j}.\label{eq:stochHamiltonian}
\end{equation}
$W_{t}^{i,j}$ describes a complex noise and we adopt the convention that $W_t^{ij *}=W_t^{j,i}$. The corresponding It\={o} rules are summed up by 
\begin{equation}\label{eq:ito}
dW_{t}^{i,j}dW_{t}^{k,l}=\delta_{i,l}\delta_{k,j}dt.
\end{equation}
Using the It\={o} rules to average over the noise degrees of freedom one recovers the Liouvillian (\ref{eq:Liouvillian}).

Finally, an other point we would like to emphasize concerns the connection to systems evolving under continuous
measurements. Indeed,  another way to unravel (\ref{eq:Liouvillian})
is to see it as the average evolution with respect to the measurement
outcomes of a system for which the variables $c_{j}^{\dagger}c_{i}+c_{i}^{\dagger}c_{j}$
and $i(c_{i}^{\dagger}c_{j}-c_{j}^{\dagger}c_{i})$
are continuously monitored and independently measured with rate $\gamma_{i,j}$ \citep{Dalibard_Unraveling}. Although the physics is radically different at the level of
a single realisation of the noise, on average it gives the same result
than the prescription (\ref{eq:stochHamiltonian}). Hence, all the results that will be presented for the mean behavior of our class of stochastic Hamiltonians
also describe the mean behavior of systems subject to continuous measurements. The unraveling procedure corresponding to  continuous measurements  is described in detail in Appendix~\ref{app:measurement}.

\subsection{Self-energy}

\begin{figure}
\begin{center}
\includegraphics[width=0.95\columnwidth]{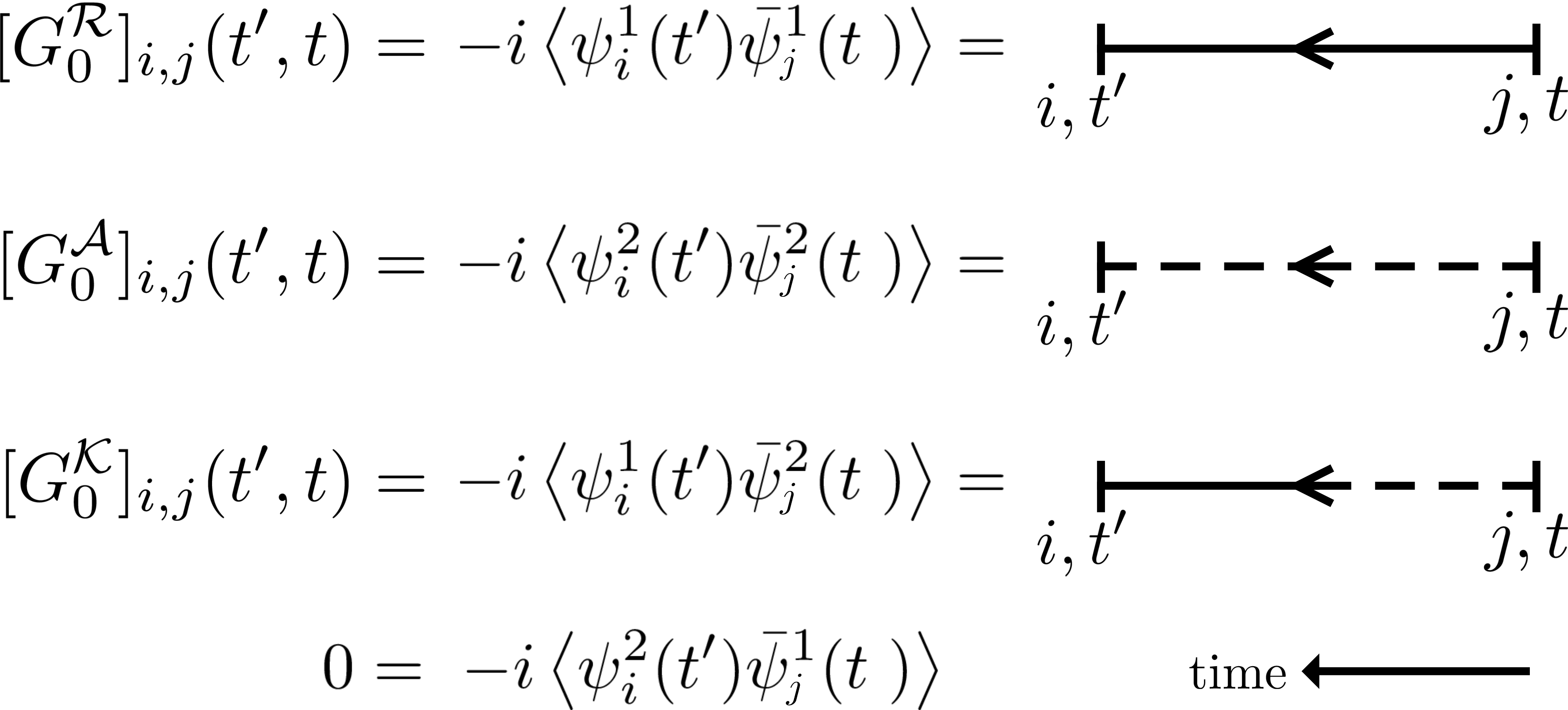}
\end{center}
\caption{\label{fig:diag-deph} 
Diagrammatic representation of the retarded ($\cal R$), advanced ($\cal A$) and Keldysh ($\cal K$) Green's function. Time flows from right to left.}
\end{figure}

We show now that the perturbation theory in the stochastic Hamiltonian~\eqref{eq:stochHamiltonian} can be fully resummed, leading to exact results for single particle Green's functions. To perform this task, we rely on the Keldysh path-integral formalism~\cite{Kamenev2011}, which describes the dynamics of the system through its action $S$. The presence of dissipative effects can be naturally included in $S$ using Lindblad formalism~\citep{Diehl_KeldyshLindblad}. The action gives the Keldysh partition function ${\cal Z}={\rm tr}(\rho_{t})$
\begin{equation}
\mathcal{Z}=\int{\cal D}[\psi^{\pm},\bar{\psi}^{\pm}]e^{iS[\psi^{\pm},\bar{\psi}^{\pm}]}.
\end{equation} 
where $\psi=(\psi^{+},\psi^{-})$ are Grassmann variables defined respectively on the positive and negative Keldysh time contours $\mathcal C_\pm$. We follow the Larkin-Ovchinnikov's convention~\footnote{In our conventions,   Larkin-Ovchinnikov's rotation reads $\psi^{1/2}=(\psi^{+}\pm\psi^{-})/\sqrt2\,, \bar{\psi}^{1/2}=(\bar{\psi}^{+}\mp\bar{\psi}^{-})/\sqrt2$~\citep{Larkin_vortices_supra}.},  in which the Keldysh action $S_0$ corresponding to the free-evolution $\mathcal L_0$
is expressed in terms of the  inverse Green's function $\boldsymbol{G}^{-1}$  namely
\begin{equation}\label{eq:S0}
\mathcal S_0=\sum_{i,j}\int\frac{d\omega}{2\pi}\left(\begin{array}{cc}
\bar{\psi}^{1}, & \bar{\psi}^{2}\end{array}\right)_i
\Big[\boldsymbol{G}^{-1}\Big]_{i,j}\left(\begin{array}{c}
\psi^{1}\\
\psi^{2}
\end{array}\right)_j\,.
\end{equation}
All variables in the integral~\eqref{eq:S0} are implicitly assumed to depend on a single frequency $\omega$, which coincides with the assumption of stationary behavior, valid for our class of problems.
The  inverse  Green's function $\boldsymbol{G}^{-1}$ is itself expressed in terms of the retarded,  advanced  and  Keldysh  green  functions $G^{\cal R/A/K}$, defined in Section~\ref{sec:Analy-D}:
\begin{equation}
\Big[\boldsymbol{G}^{-1}\Big]=\left(\begin{array}{cc}
G^{\cal R} & G^{\cal K}\\
0 & G^{\cal A}
\end{array}\right)^{-1}
\end{equation}
and whose diagrammatic representations in the time domain are given in Fig.~\ref{fig:diag-deph}. 
The causality structure of the Keldysh Green functions is enforced by the suppression of correlators $\langle\psi^{2}\bar{\psi}^{1}\rangle=0$. This means that a retarded propagator  can never become advanced, which pictorially translates into the fact that a solid line cannot switch to a dashed one. 

The action corresponding to the Liouvillian term~\eqref{eq:Liouvillian} reads \cite{Diehl_KeldyshLindblad}
\begin{equation}\label{eq:stoca}
S_{\cal L}:=-\int dt \sum_{i,j}\gamma_{i,j}\left(\bar{\psi}^1_{j,t} \psi^1_{i,t} \bar{\psi}^2_{i,t} \psi^2_{j,t} + \bar{\psi}^1_{i,t} \psi^1_{j,t} \bar{\psi}^2_{j,t} \psi^2_{i,t}\right)\,.
\end{equation}
which is a quartic action in the Grassmann fields. At the level of single particle Green's functions, the action $S_{\cal L}$ is incorporated through the self energy $\boldsymbol\Sigma$, defined as the sum of all one-particle irreducible diagrams. As
in equilibrium field theory, the Dyson equation relates the full propagator
to the bare propagator and the self energies $\boldsymbol{\Sigma}$: 
\begin{equation}\label{eq:DysonM}
\boldsymbol{G}=\Big[\boldsymbol{G}_{0}^{-1}-\boldsymbol{\Sigma}\Big]^{-1}\,.
\end{equation}
To compute the diffusive current from MW formula, $\boldsymbol{\Sigma}$ must be know to any order; an {\it a priori} difficult task given the quartic nature of the action~\eqref{eq:stoca}. Instead, rewriting the action at the stochastic level allows us to exactly derive the self-energy $\boldsymbol\Sigma$ and solve this problem. In the field-theory language, the unraveling procedure exemplified by Eq.~\eqref{eq:stochHamiltonian} leads to the equivalent action
\begin{equation} \label{eq:stochasticaction}
S_{{\rm sto}} =  -\sum_{i,j}\int\sqrt{2\gamma_{i,j}}\left(\bar{\psi}_{j,t}^{1}\psi_{i,t}^{1}+\bar{\psi}_{j,t}^{2}\psi_{i,t}^{2}\right)dW_{t}^{i,j} \,, 
\end{equation}
where $S_{\rm sto}$ is related to $S_{\cal L}$ by the average $\mathbb{E}[]$ over the noise degrees of freedom:
\begin{equation}
    \mathbb{E}[e^{iS_{\rm sto}}]=e^{iS_{\cal L}}\,.
\end{equation}
In formal terms, this transformation is reminiscent of a Hubbard-Stratonovich transformation where the  action becomes  quadratic in terms
of the Grassmann variables. Note that the complexity encoded in Eq.~\eqref{eq:stoca} is preserved by the consequent introduction of the space and time dependent noise $dW_t^{i,j}$. However, the noise correlations imposed by It\=o's rules~\eqref{eq:ito} allow a dramatic simplification of  the diagrammatic expansion
in $\gamma_{i,j}$ of the Green functions  within the stochastic formulation. Such simplified structure does not manifestly appear when working with the Lindbladian (averaged)  formulation of the problem~\eqref{eq:stoca} (see Fig.~\ref{fig:crossing} in Appendix~\ref{sec:App-Stochastic}).

\begin{figure}
\begin{center}
\includegraphics[width=0.95\columnwidth]{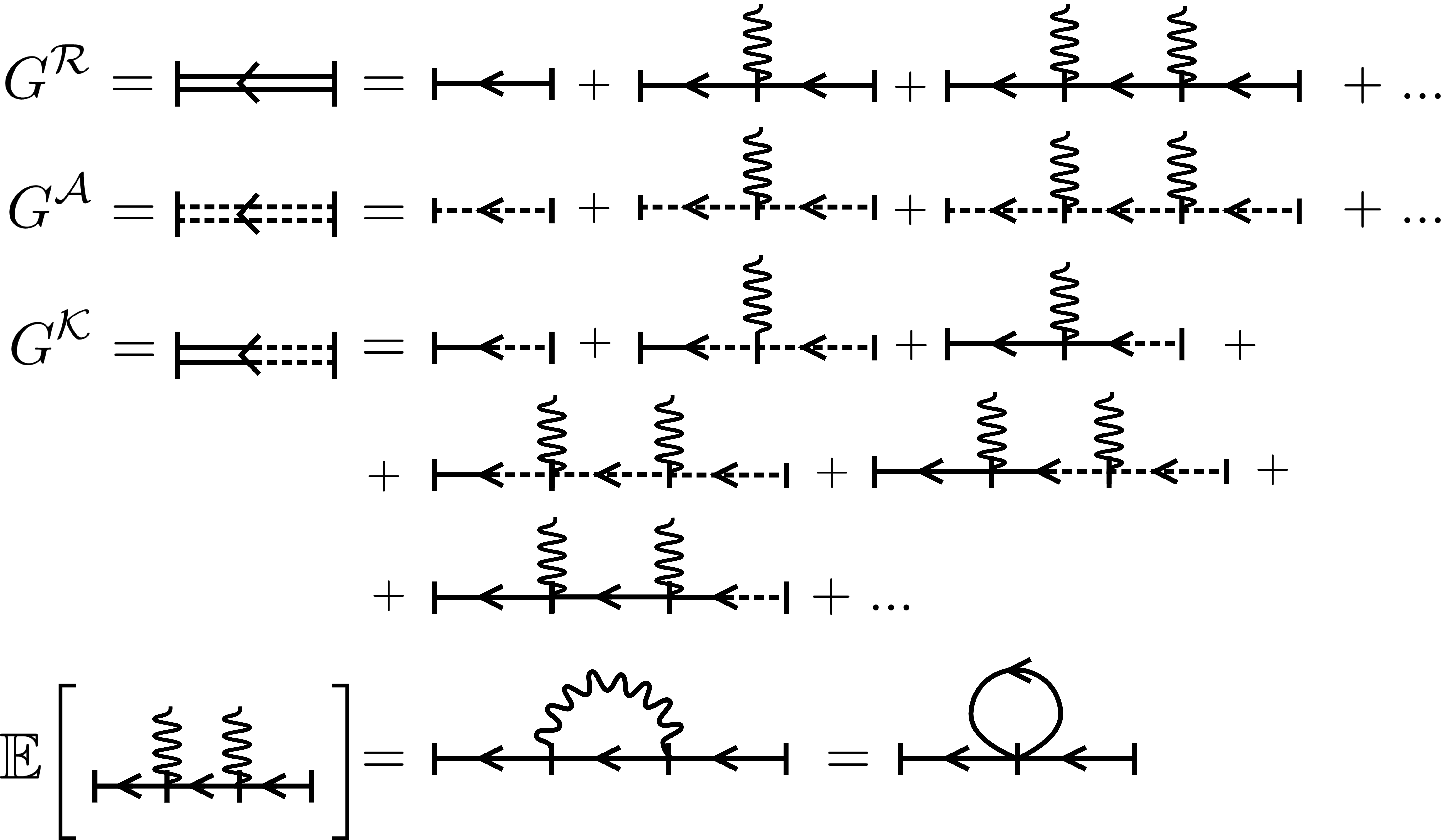}
\end{center}
\caption{\label{fig:perturbativeexpansion}
Perturbative series in the Keldysh formalism for our class of stochastic
models. Average quantities are obtained by contracting pairs of wiggly
lines together. Here a wiggly line represents either $dW_{t}^{i,j}$
or its complex conjugated pair for simplicity. The formulation of the theory in terms of QSH allows for a simple writing of the perturbative expansion.}
\end{figure}

The resummation works as follows. In Fig.~\ref{fig:perturbativeexpansion}, we show the diagrammatic expansion of \eqref{eq:DysonM} up to second order in the stochastic noise $\gamma_{i,j}$. 
The wiggly lines represent $dW_{t}^{i,j}$.
Since we are interested in the mean behavior, we have to take the
average over the noise degrees of freedom. This amounts to contract
wiggly lines pair by pair. From the It\={o} rules~\eqref{eq:ito}, we see that upon contraction,
a wiggly line forces the two vertices it connects to have the same time and position, as illustrated in  Fig.~\ref{fig:perturbativeexpansion}.

\begin{figure}
\begin{center}
\includegraphics[width=0.95\columnwidth]{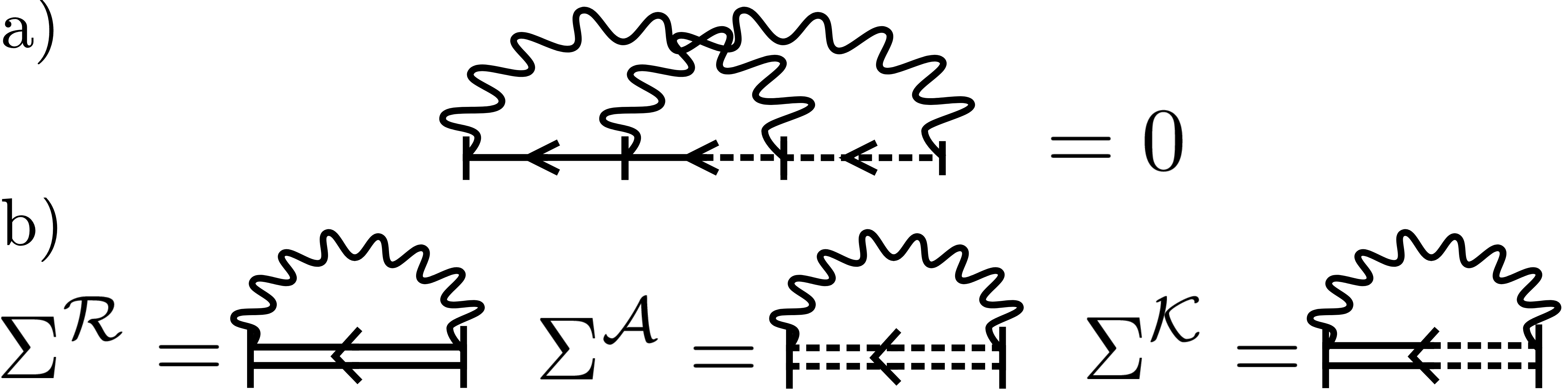}
\end{center}
\caption{\label{fig:scba}
a) Non-crossing rule for the contraction of wiggly lines. b) Self-energies
for the different Keldysh components.}
\end{figure}

The important consequence is that all the diagrams which present a \emph{crossing }of the wiggly lines vanish because of the
causal structure of the Keldysh's Green function, namely that $G^{\cal R}(t,t')$
is non zero only for $t>t'$ and conversely for $G^{\cal A}$. For a detailed proof
of this statement, see Appendix~\ref{sec:App-Stochastic}. In particular, the constraints of avoided wiggly lines  establishes the validity of the self-consistent Born approximation (SCBA) for 
the self-energy of single particle Green's function and generalize the approach presented in Ref.~\cite{dolgirevNonGaussianCorrelationsImprinted2020a}. SCBA allows a simple and compact
derivation of all components as exemplified by the diagrammatic representation in Fig.~\ref{fig:scba}. 
Namely, we have that in position space 
\begin{equation}
\boldsymbol{\Sigma}_{i,j}(t,t')=\delta_{i,j}\delta(t,t')\sum_{k}\gamma_{i,k}\boldsymbol{G}_{k,k}(t,t)\,. \label{eq:self}
\end{equation}
For the retarded and advanced components, this relation takes a particularly
simple form since $G_{j,k}^{\cal{R}(\cal{A})}(t,t)=\mp\frac{i}{2}\delta_{j,k}$ in position space. Note that this simple expression is only valid when the two time indices are taken to be equal and comes entirely from the causal structure of the Green's functions in the Keldysh formalism. One way to see this is to evaluate the step function $\theta(t-t')$ for the retarded and advanced Green's functions from the discrete version of the path integral presented in 9.2 of [80].
To get the Keldysh component $G^{\cal K}$, one has to solve the self-consistent Dyson equation : 
\begin{equation}
\boldsymbol G^{\cal K}=-\boldsymbol G^{\cal R}\left(\left[\boldsymbol G_{0}^{-1}\right]^{\cal K}-\boldsymbol \Sigma^{\cal K}\right)\boldsymbol G^{\cal A}\,, \label{eq:Dyson}
\end{equation}
which is a problem whose complexity only scales polynomially with the number of degrees of freedom in the system (such as the system size $N$ of the setup in Fig.~\ref{fig:Generic-setup}). This solves the problem entirely at the level of single-particle  correlation
functions. Remark that this applies to any model as long as the bare theory respects a Wick's theorem and its propagators are known. It allows a systematic study of quantum systems in the presence of external noisy degrees of freedom. 

This ability to calculate the Keldysh  Green's function is crucial to give an exact description of out-of-equilibrium transport  in dissipative systems, as we are going to show in the next section.

\section{Applications}\label{sec:applications}

We now proceed to employ the self-consistent approach to showcase our $1/N$ expansion, presented in Sec.~\ref{sec:Analy-D}, against a large class of QSHs that display diffusive transport.
 
The action describing the out-of-equilibrium setting represented in Fig.~\ref{fig:Generic-setup} has the form
\begin{equation}
S=S_{{\rm Bd}}+S_{0}+S_{{\rm sto}}.
\end{equation}
The first term in the action, $S_{{\rm Bd}}$,
describes the exchange coupling with gapless non-interacting fermionic reservoirs of chemical potential $\mu_{L,R}$ and temperature $T_{L,R}$. The corresponding action, under the assumptions discussed in Section~\ref{sec:Analy-D},  was derived for instance in Ref.~\citep{JinFilipponeGiamarchi_GenericMarkovian}: 
\begin{align}
S_{{\rm Bd}} & =i\Gamma\sum_{a=L,R}\int\frac{d\omega}{2\pi}\bar{\psi}_{a}\begin{bmatrix}1 & 2\tanh\left(\frac{\omega-\mu_{a}}{2T_{a}}\right)\\
0 & -1
\end{bmatrix}\psi_a\,,
\end{align}
where $\psi_a$ is a shorthand notation for $(\psi_a^1,\psi_a^2)$, $L$ designates site 1 and $R$ designates site $N$. The action $S_0$ is the quadratic action related to the intrinsic dynamics of the system, which can describe various situations from coherent dynamics to  single-particle dissipative gains and losses~\cite{JinFilipponeGiamarchi_GenericMarkovian}.  In this paper, we will focus  on one-dimensional nearest-neighbour coherent bulk hopping, which is described by the standard Hamiltonian,   \begin{equation}
H_{\tau}:=\tau\sum_{j=1}^{N-1}\left(c_{j}^{\dagger}c_{j+1}+c_{j+1}^{\dagger}c_{j}\right)\label{eq:coherentjumpHamiltonian},
\end{equation}
with $\tau$ the hopping amplitude. The corresponding action reads
\begin{equation}
S_0=-i\tau\sum_{j}\int\frac{d\omega}{2\pi}\Big(\bar{\psi}_{j}^{1}\psi_{j+1}^{1}+\bar{\psi}_{j}^{2}\psi_{j+1}^{2}+\mbox{c.c}\Big)\,.\label{eq:actiondyn}
\end{equation}
The free propagators are directly derived from  the previous expressions of the action and read
\begin{align}
\begin{split}
\left[G_0^{-1}\right]_{j,k}^{\cal{R}(\cal{A})}(\omega)= & \delta_{j,k}\Big[\omega\pm i\Gamma(\delta_{j,1}+\delta_{j,N})\Big]\label{eq:bareprop}\\
 &\qquad\qquad+\tau(\delta_{j,k+1}+\delta_{j,k-1})\,,
 \end{split}
 \\
\left[G_0^{-1}\right]_{j,k}^{\cal K}(\omega)= & 2i\Gamma\delta_{j,k}\sum_{a=L,R}\delta_{j,a}\tanh\left(\frac{\omega-\mu_{a}}{2T_{a}}\right)\,. \label{eq:bareprop2}
\end{align}
Notice that the reservoirs act, through the hybridization constant $\Gamma$, as  natural regulators of the imaginary components of the non-interacting problem~\cite{Kamenev2011}.

Finally $S_{{\rm sto}}$ is the action corresponding
to the QSH (\ref{eq:stochasticaction}).  
As explained in the previous section, the demonstrated validity of SCBA for the Dyson
equation~\eqref{eq:Dyson} allows to derive exact expressions for the self-energies~\eqref{eq:self},  and thus for  the propagators of the full theory. Such solution allows to fully determine the transport properties of the system through MW formula~\eqref{eq:MWDn}. 
As shown in Section~\ref{sec:SCBA}, Equation~\eqref{eq:self} implies a particularly simple form for the advanced and retarded components of  the self-energy:
\begin{equation}\label{eq:selfRA}
\Sigma^{\cal R(A)}_{i,j}=\mp i\delta_{i,j}\delta(t,t')\sum_{l}\frac{\gamma_{i,l}}{2	}\,.
\end{equation}
Importantly, in the geometry of Fig.~\ref{fig:Generic-setup}, we can derive a compact and explicit expression of \eqref{eq:Dyson} for the diagonal terms $G^{\cal K}(t,t)$
\begin{equation}
\vec{G}^{\cal K}=(\mathbb{I}-M)^{-1}\cdot\vec{V}\label{eq:GK}
\end{equation}
where we introduced the $N$-dimensional vectors 
\begin{align}
\vec{G}_j^{\cal K}&=G_{j,j}^{\cal K}(t,t)\,,\\
\vec{V}_{j}&=\frac{2\Gamma}i\sum_{a\in\{L,R\}}\int\frac{d\omega}{2\pi}G_{j,a}^{\cal R}G_{a,j}^{\cal A}\tanh\left(\frac{\omega-\mu_{a}}{2T_a}\right)\,,\label{eq:V}
\end{align}
and $M$ is an $N\times N$ matrix with elements
\begin{equation}\label{eq:M}
M_{j,k}=\sum_{l}\gamma_{k,l}\int\frac{d\omega}{2\pi}G_{j,l}^{R}G_{l,j}^{A}\,.
\end{equation}
Notice that  only $G^{\mathcal K}$ carries information about the biased reservoirs, as can be seen from \eqref{eq:V}. 
The first term in \eqref{eq:MWDn} depends exclusively on spectral functions, which are readily derived from Eqs.~\eqref{eq:bareprop} and~\eqref{eq:selfRA}, while Eq.~\eqref{eq:GK} sets, through Eq.~\eqref{eq:GKexpansion}, the expression of the density differences at the edges $\Delta n$.

Note that our analysis shows that the matrix $M$~\eqref{eq:M} is the key object encoding information about diffusion and it appears exclusively in the Keldysh component of the single-particle Green's function~\eqref{eq:GK}. A convenient way to understand this is to consider systems with single-particle gains and losses that do not display Ohmic $1/N$ suppression of the current. It was shown in Ref.~\cite{JinFilipponeGiamarchi_GenericMarkovian} that, while \eqref{eq:selfRA} remains valid in those systems, the matrix $M$ in \eqref{eq:GK} becomes 0 for these systems and the current saturates to a size-independent value. Thus, having a finite-lifetime in the retarded and advanced Green's function is not sufficient to get diffusive transport.
The imaginary contribution to the retarded/advanced self-energy, such as the one in \eqref{eq:selfRA}, has the interpretation of a lifetime for the free single-particle excitations of the system, yet it is the Keldysh component of the self-energy that describes the consequences of dissipative scattering on the transport properties of the system. When $M\neq 0$, equation \eqref{eq:M} gives us a linear profile for the density profile, which eventually leads to a $1/N$ diffusive contribution for the current as discussed in the \ref{sec:Analy-D}.
 
\begin{figure}
\begin{center}
\includegraphics[width=0.8\columnwidth]{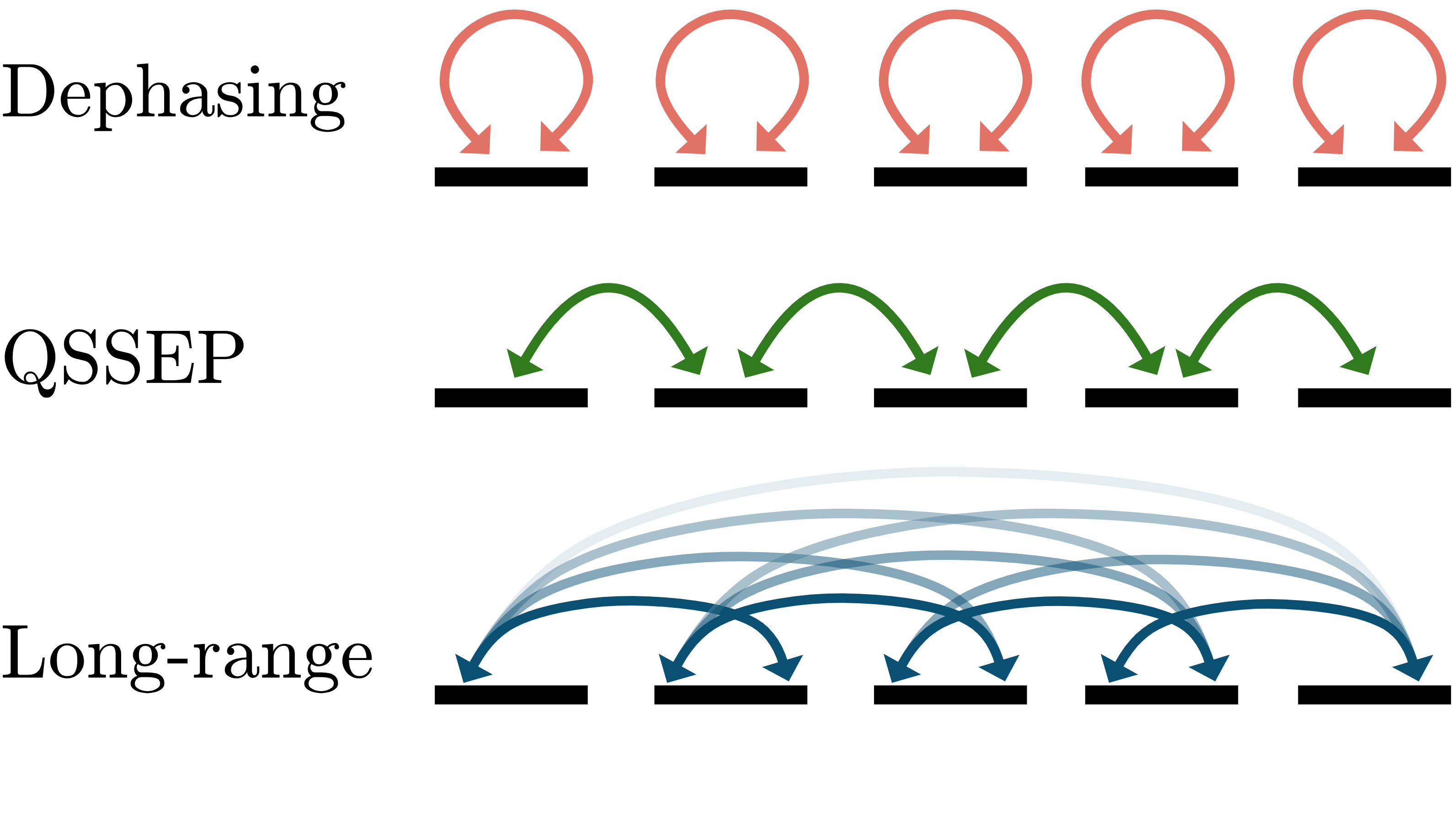}
\end{center}
\caption{\label{fig:differentmodels}
Particular 1D discrete cases that will be of interest. Only the noise contribution is presented in this figure. In the dephasing model, all the sites are paired with themselves. For the QSSEP, the
pairs are between nearest neighbours. In the long-range model, a given point is linked to all the rest of the lattice with a coupling decaying as power-law.}
\end{figure}

These considerations are those underpinning our general  discussion about diffusive transport in Sec.~\ref{sec:Analy-D}.  We now turn to the case-by-case study of the specific QSHs depicted in Fig.~\ref{fig:differentmodels}. As said in the Introduction, we will focus on three one-dimensional models: the dephasing model, the quantum
symmetric simple exclusion process (QSSEP) and models with stochastic long-range hopping. For the dephasing model, every single point on the lattice
is coupled with itself by the noise. For the QSSEP, the noise couples each point with its  neighbours. For the long-range, a given point is paired to all the rest of the lattice with a power-law decay as a function of the distance. These processes are illustrated for all three models in Fig.~\ref{fig:differentmodels} and we will give more details about their physical motivations in the related sections.

Without loss of generality, in the oncoming analysis of the current $J$, we focus on a linear response regime in the chemical potential bias. We set an idential temperature for both reservoirs $T_{L}=T_{R}=T$ and $\mu_{L}\to\mu+\delta\mu,\hspace{1em}\mu_{R}\to\mu-\delta\mu$.
We expand Eq.~\eqref{eq:MWDn} in $\delta\mu$. One thus obtains, to linear order in $\delta\mu$:
\begin{equation}\label{eq:linearreponse}
J=\Gamma\frac{\delta\mu}{2T}\int d\omega\frac{1}{\cosh^{2}\left(\frac{\omega-\mu}{2T}\right)} \left[\mathcal A(\omega)-\frac\Gamma{2\pi}\Delta^{\mathcal K}(\omega)\right]\,.
\end{equation}
where $\mathcal A(\omega)$ is the edge spectral function, which coincides with $\mathcal A_{L/R}(\omega)$, because of the mirror symmetry of the class of QSHs that we will consider. The second term  can be expressed in the form 
\begin{equation}\label{eq:GKvec}
    \Delta^{\mathcal K}(\omega)=\left[\frac1{\mathbb{I}-M}\cdot\vec W(\omega)\right]_1-\left[\frac1{\mathbb{I}-M}\cdot\vec W(\omega)\right]_N\,,
\end{equation}
in which  $\vec W$ is an $N$ dimensional vector
whose components are given by $\vec W_j(\omega)=G_{j,1}^{\cal R}(\omega)G_{1,j}^{\cal A}(\omega)-G_{j,N}^{\cal R}(\omega)G_{N,j}^{\cal A}(\omega)$.

\subsection{Dephasing model} \label{sec:DPH}

The dephasing model describes fermions hopping
on a 1D lattice while subject to a random onsite dephasing coming from dissipative
interactions with external degrees of freedom. In the language of Sec.~\ref{sec:SCBA}, this model corresponds to the case where all the points are paired with themselves, which results in substituting the rates 
\begin{equation}\label{eq:gammadph}
    \gamma_{i,j}\rightarrow\, \gamma_{\rm Dph}\delta_{i,j}\,,
\end{equation}
in Eqs.~\eqref{eq:Liouvillian} and~\eqref{eq:stochHamiltonian}  (see also Fig.~\ref{fig:differentmodels}). There are various limits
in which this model can be derived. For instance, it can be thought as describing the effective dynamics of fermions interacting weakly with
external bosonic degrees of freedom within the Born-Markov approximation \citep{BreuerPetruccione_book}. In Refs.~\citep{Znidaric__XXdeph,Znidaric_dephasing,Znidaric__dephasingXXZ} it was shown, relying on  matrix product operator techniques, that the dephasing model exhibits diffusive transport. Two-times correlators in the XXZ under dephasing was also studied in \citep{wolffEvolutionTwotimeCorrelations2019} and were shown to exhibit a complex relaxation scheme. For bosonic interacting systems, it was shown that the addition of an external dephasing could lead to anomalous transport \citep{Kollath_Anomaloustransportbosons,Kollath_Glasslikebosons}. Additionally, as discussed in Section~\ref{sec:SCBA}, the mean dynamics of this model coincides with the one where the occupation numbers of fermions on each site are independently and continuously monitored~\citep{BernardJinShpielberg_2018,caoEntanglementFermionChain2019}. For this reason, the dephasing model has recently attracted a lot of interest as a prototypical model exhibiting a measurement rate-induced transition in the entanglement dynamics~\citep{albertonTrajectoryDependentEntanglement2020,buchholdEffectiveTheoryMeasurementInduced2021}. Finally, we note that in Ref.~\citep{ProsenEssler_Mapping} a mapping between the dephasing
model and the Fermi-Hubbard model was established. Although we will not discuss this mapping here, we stress that it implies that our method also provides the computation of exact quantities valid for equivalent systems governed by Hubbard Hamiltonians. 

The stochastic Hamiltonian for the dephasing model is readily obtained from the substitution~\eqref{eq:gammadph}, namely  
\begin{equation}\label{eq:dh_dephasing}
dH_{t}=\sqrt{2\gamma_{{\rm Dph}}}\sum_{j}\hat{n}_{j}dB_{t}^{j}\,,
\end{equation}
where $B_{t}$ denotes a real Brownian motion with It\={o} rule $dB_t^j dB_t^k=\delta_{j,k}dt$.  The retarded and advanced self-energies are obtained from Eq.~\eqref{eq:selfRA} and read
\begin{equation}\label{eq:selfdph}
\Sigma_{j,k}^{\cal{R}(\cal{A})}(t,t')=\mp\frac{i}{2}\gamma_{{\rm Dph}}\delta_{j,k}\delta(t-t')\,.
\end{equation}
while $G^{\mathcal{R},\mathcal{A}}$ are obtained by inversion of Eq.~\eqref{eq:bareprop} with inclusion of the self-energy~\eqref{eq:selfdph}. These functions are symmetric and given
by, for $i\leq j$~\citep{Tan_InverseTridiagonal,JinFilipponeGiamarchi_GenericMarkovian}: 
\begin{equation}\label{eq:GRdeph}
G_{i,j}^{\cal{R}/\cal{A}}(\omega)=\frac{(-1)^{i+j}\tau^{j-i}B_{i-1}^{\cal{R}/\cal{A}}B_{N-j}^{\cal{R}/\cal{A}}}{\left[\omega\pm i\left(\Gamma+\frac{\gamma_{{\rm Dph}}}{2}\right)\right]B_{N-1}^{\cal{R}/\cal{A}}-\tau^{2}B_{N-2}^{\cal{R}/\cal{A}}}\,,
\end{equation}
where $B_{i}^{\cal{R}/\cal{A}}=[(r_{+}\pm i\Gamma)r^i_+-(r_{-}\pm i\Gamma)r_{-}^{i}]/(r_{+}-r_{-})$ and 
$r_{\pm}=\left(\omega\pm i\frac{\gamma}{2}+\sqrt{(\omega\pm i\frac{\gamma_{{\rm Dph}}}{2})^2-4\tau^{2}}\right)/2$.

\begin{figure}
\begin{center}
\includegraphics[width=\columnwidth]{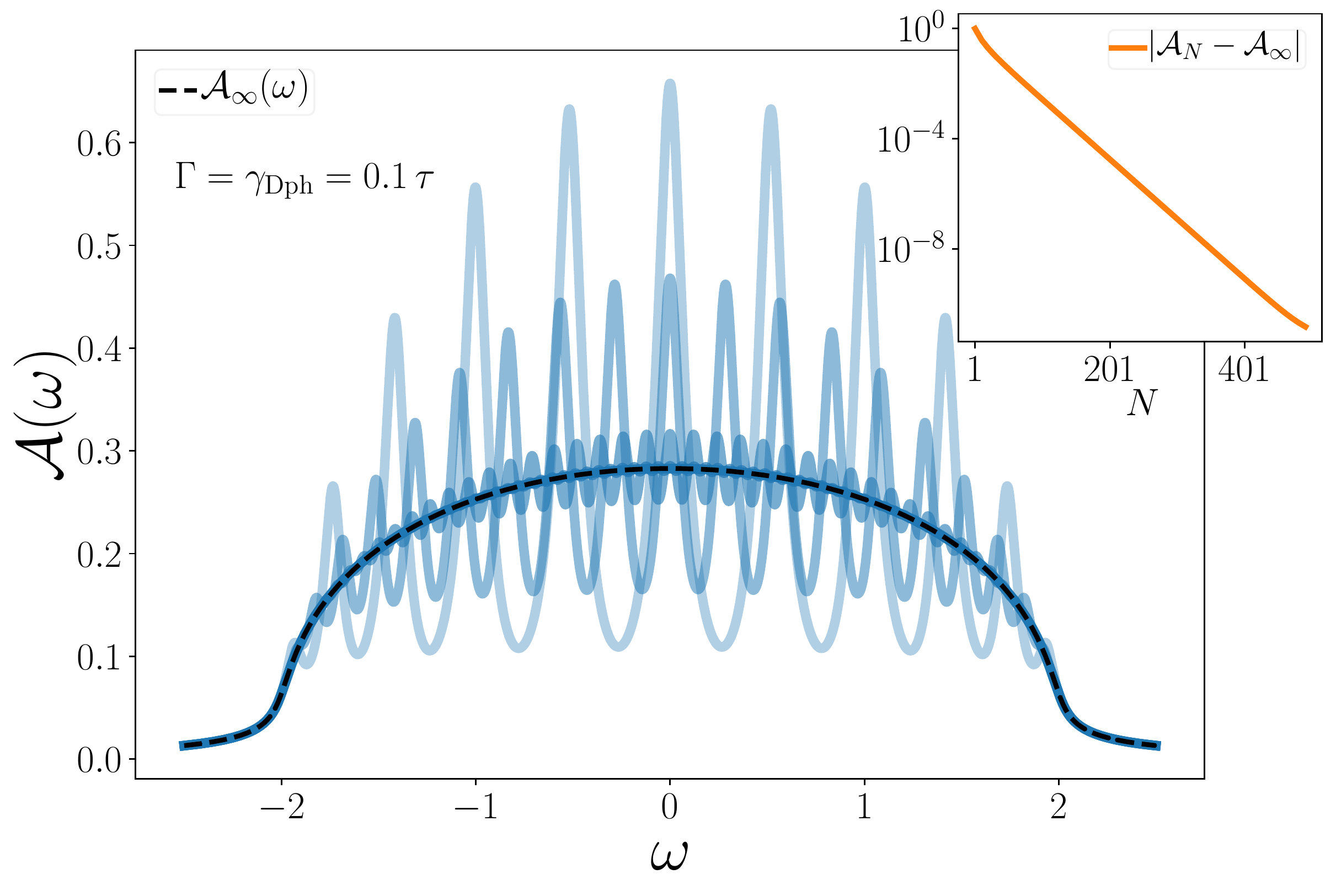}
\end{center}
\caption{\label{fig:spectral_dephasing}
Edge spectral function $\mathcal A(\omega)$ for the dephasing model~\eqref{eq:dh_dephasing} in the configuration of Fig.~\ref{fig:Generic-setup} for difference systems sizes $N$. Darker blue solid lines correspond to larger systems sizes $N=11,21,51,101,201,501,1001$. We consider only odd values of $N$, as they ensure the presence of a resonance at $\omega=0$. The inset shows the exponential convergence of the spectral function at a fixed (odd) system size $\mathcal A_N=\mathcal A(\omega=0)$ towards its asymptotic value $\mathcal A_\infty(\omega)$, obtained from Eq.~\eqref{eq:GRinf} and corresponding to the dashed black line in the main plot (for $N\gtrsim100$ and the parameters reported in the plot, numerical curves overlap with  $\mathcal A_\infty(\omega)$).}
\end{figure}

The related spectral functions at the system edges $\mathcal A(\omega)=\mathcal A_{11}(\omega)=\mathcal A_{NN}(\omega)$ is represented in Fig.~\ref{fig:spectral_dephasing} for different system sizes $N$. 
It displays $N$ peaks corresponding to the eigenspectrum of the system without dissipation. The width of the peaks is controlled non-trivially by the hybridization constant $\Gamma$ and the bulk dissipation rate $\gamma_{\rm Dph}$. Plots for closely related  quantities in the $\gamma_{\rm Dph}\rightarrow0$ limit can be found in Ref.~\cite{JinFilipponeGiamarchi_GenericMarkovian}. In this nondissipative limit, the height of the peaks does not decay with the system size $N$. On the contrary, for  $\gamma_{\rm Dph}>0$, the peaks vanish in the $N\rightarrow\infty$ limit, and the spectral function converges {\it exponentially} towards a smooth function  $\mathcal A_\infty(\omega)$ as shown in the  inset of Fig.~\ref{fig:spectral_dephasing}. One can analytically derive $\mathcal A_\infty(\omega)$, as the retarded Green function~\eqref{eq:GRdeph} at the edges $G^{\cal R}_{1,1}=G^{\cal R}_{N,N}$ converges to
\begin{equation}\label{eq:GRinf}
    \lim_{N\ra \infty}G^{\cal R}_{1,1}(\omega)=\frac1{\omega+i\left(\Gamma+\frac{\gamma_{\rm Dph}}2\right)-\frac{\tau^2}{r_{\rm sgn(\omega)}}}\,.
\end{equation}

The exponential convergence of the edge spectral function is reproduced by all the other QSHs discussed below and verifies one of the preliminary assumptions exposed in  Section~\ref{sec:Analy-D}, identifying  the density difference $\Delta n$ as the term entirely responsible for  the $1/N$ suppression of the dissipative current in \eqref{eq:MWDn}. 

Our approach provides an efficient way to compute the second term in \eqref{eq:linearreponse}, through an explicit derivation of the matrix $M$:
\begin{equation}\label{eq:Mdephasing}
M_{j,k}=\gamma_{\rm Dph }\int\frac{d\omega}{2\pi}G_{j,k}^{\cal R}G_{k,j}^{\cal A}\,.
\end{equation}

As we detail in Appendix~\ref{sec:App-Numerics}, the expressions~\eqref{eq:GKvec},~\eqref{eq:GRdeph} and~\eqref{eq:Mdephasing} allow the efficient derivation of the current~\eqref{eq:linearreponse} up to system sizes $N\simeq10^{3-4}$. As a consequence, we can systematically study the expected crossover from a ballistic-to-diffusive regime expected at length scales $N^*\simeq\gamma_{{\rm Dph}}^{-1}$~\citep{Znidaric__XXdeph}. See also  Appendix~\ref{sec:App-Numerics_2} for additional details.

\begin{figure}
\begin{center}
\includegraphics[width=0.95\columnwidth]{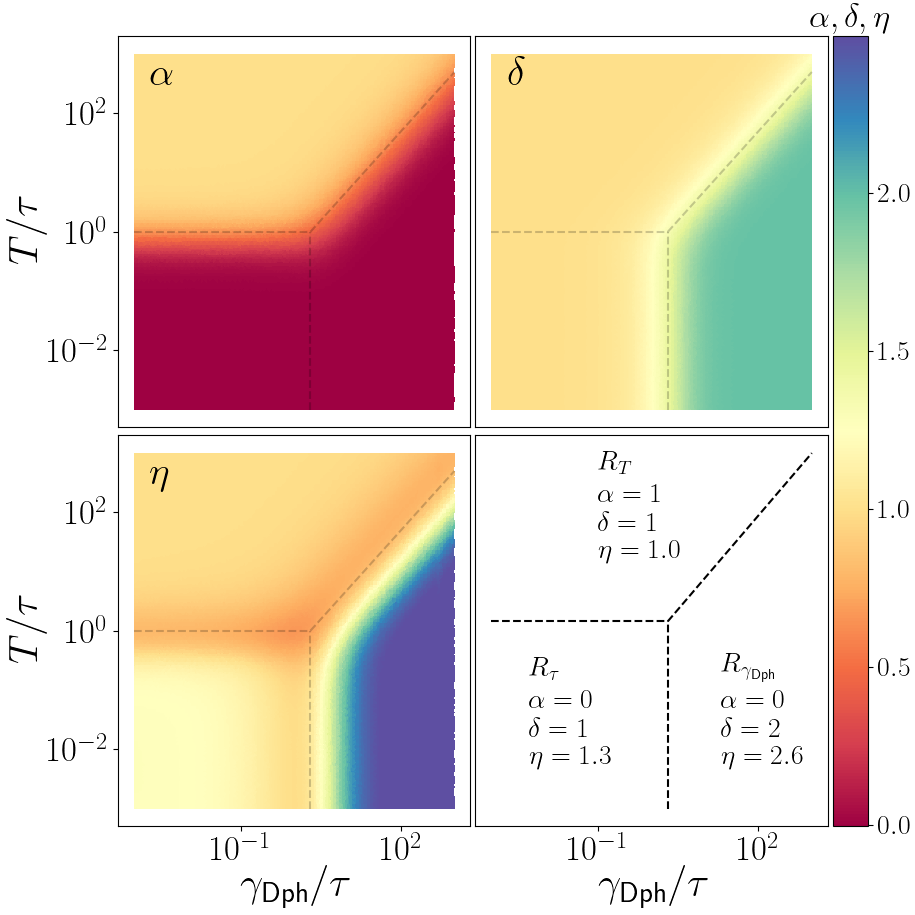}
\end{center}
\caption{\label{fig:current_phase_diagram-1}
Fitted parameters $(\alpha,\delta,\nu)$ of the rescaled conductance of the  dephasing model as defined in Eq.~\eqref{eq:Jdeph}.  These values define different regions in the temperature - dephasing plane with different behaviors for the conductance, see Eq.~\eqref{eq:JTdeph}. The dashed lines are a guide for the eyes to delimit the regions. The bottom right plot summarizes the characteristic values of each region.}
\end{figure}

Two main technical advances of our approach compared to previous studies \citep{Znidaric__XXdeph,Znidaric__dephasingXXZ,KarevskiRepeatedinterac,yamanaka2021exact,Guo2017,Znidaric_dephasing,TurkeshiSchiro_Dephmodel} consist in its ability to naturally address reservoirs with finite temperatures $T<\infty$, accessing transport regimes left unexplored by previous studies and to access two-times correlators in the stationary state. An important consequence of our analysis is that the rescaled  conductance  of the system, that we define as $\mathcal G=NJ/\delta\mu$, has a non-trivial dependence on the temperature $T$ and the dephasing rate $\gamma_{\rm Dph}$, namely
\begin{equation}\label{eq:Jdeph}
\mathcal G=\lim_{N\rightarrow\infty}\frac{JN}{\delta\mu}=\frac{\eta \tau^{\alpha+\delta}}{T^{\alpha}\gamma_{{\rm Dph}}^{\delta}}\,.
\end{equation}
In Fig.~\ref{fig:current_phase_diagram-1}, we plot the coefficients
$(\a,\d,\eta)$ across the parameter space $(T,\gamma_{{\rm Dph}})$.
From the plot, we identify three main diffusive transport regimes, $R_{\tau,T,\gamma}$, in which these coefficients are different.
Note that the regions are not connected
by sharp phase transitions but instead by crossovers, which appear
sharp in logarithmic scale. Deep in the three regions, the rescaled conductance
takes the approximate values
\begin{equation}\label{eq:JTdeph}
\mathcal G=\begin{cases}
\frac{\tau^2}{T\gamma_{{\rm Dph}}} &T\gg\gamma_{{\rm Dph}},\t\\
\frac{2.6\tau^2}{\gamma_{{\rm Dph}}^2} & \gamma_{{\rm Dph}}\gg T,\t\\
\frac{1.3\tau}{\gamma_{{\rm Dph}}} & \t\gg\gamma_{{\rm Dph}},T 
\end{cases}\,.
\end{equation}

In previous studies carried in the $T\rightarrow\infty$ limit for the reservoirs, where they can be described as Lindblad injectors~\cite{JinFilipponeGiamarchi_GenericMarkovian}, the conductance $\mathcal G$ is assumed to be proportional to the bulk diffusion constant $D$~\cite{bertinifinite2021,znidaricNonequilibriumSteadystateKubo2019a}. The density profiles in the system (see App.~\ref{sec:App-Numerics_2}) clearly show that such interpretation cannot be extended  to lower temperatures. The emergence of coherent effects between the system and its baths leads to  finite-sized boundary effects, which do not allow the determination of the bulk diffusion constant through Eq.~\eqref{eq:JTdeph}. To obtain the bulk diffusion constant we can use our approach to derive the density profiles \emph{inside the system} and far away from its boundaries. We numerically verify Fick's law~\eqref{eq:fick} in the bulk and find the diffusion constant to be
\begin{equation}\label{eq:ddeph}
D=\frac{2\tau^2}{\gamma_{\rm Dph}}\,,
\end{equation}
which is double the conductance in the $T\gg\gamma_{{\rm Dph}}$ limit, as expected.
At variance with the rescaled conductances~\eqref{eq:JTdeph}, this quantity is not affected by any boundary effect and it is in agreement with previous analytical ansatzes, valid in the infinite temperature limit~\cite{Znidaric__XXdeph}. The independence of the diffusion constant~\eqref{eq:ddeph} from the temperature at the boundaries is a consequence of the stochastic dephasing~\eqref{eq:dh_dephasing}, which locally brings the system back to an infinite temperature equilibrium state regardless of boundary conditions.  
We thus see on this example that our approach allows to compute both the two- and four-points measurements of the resistance. Even for diffusive systems, the distinction between the two processes can be important. 

To conclude our analysis of the transport in the dephasing model, we note that the different transport regimes in \eqref{eq:JTdeph} explicitly depend on the \emph{stationary
bias} $n_{1}-n_{N}$, which suffers from boundary effects  in some regions of the $(T,\gamma_{{\rm Dph}})$ parameter space. We confirm
with our exact numerical solution that this is indeed the case. This interesting bias dependence is beyond the scope of the present paper and left for future studies. 

\subsubsection*{1/N expansion}

Let us now show how the diffusion constant~\eqref{eq:ddeph}, that we obtained from our exact solution, can also be easily derived from the novel $1/N$ perturbative theory we introduced in Sec.~\ref{sec:Analy-D}. 

\begin{figure*}
\begin{center}
\includegraphics[width=0.98\textwidth]{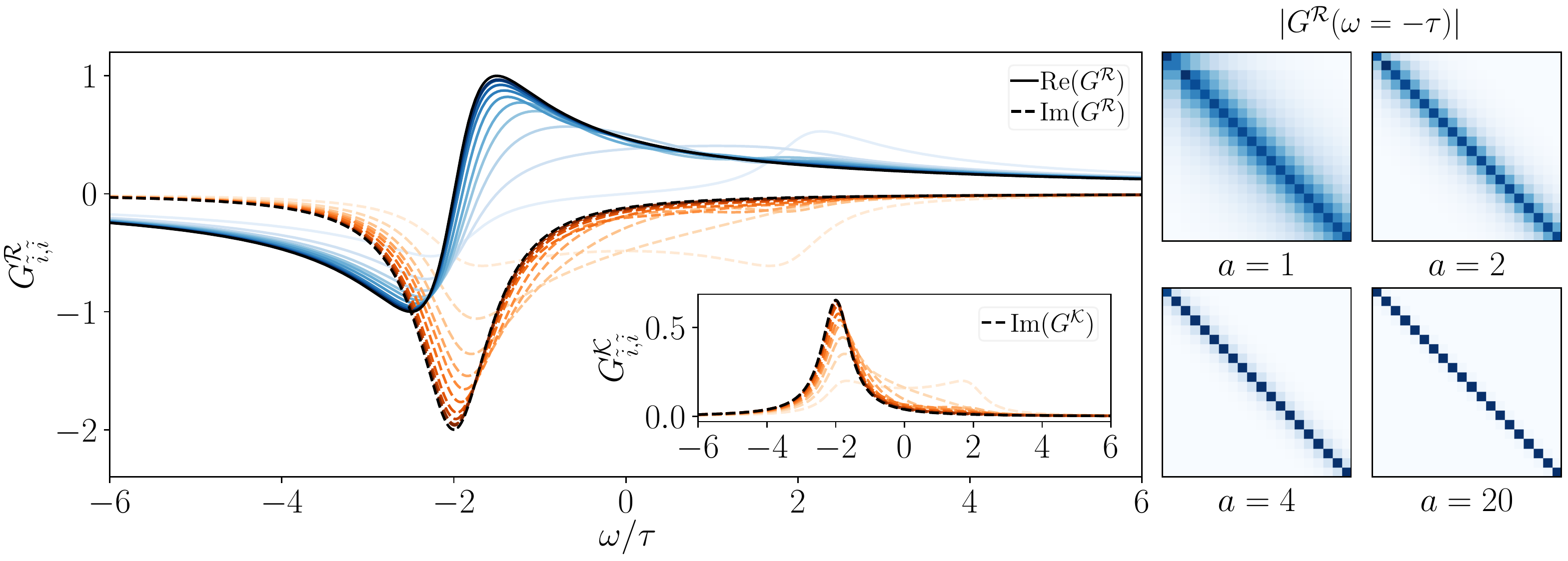}
\end{center}
\caption{\label{fig:RG} 
Coarse-graining procedure in the dephasing model, $\gamma_{\text{Dph}}=1$ for increasing size of the cell, $a$. Left: Real and imaginary part of the diagonal terms of $G^{\cal R}(\omega)$  for increasing cell size, $a=1,2,3,4,5,7,12,20,40,50$, respectively from light to dark. Inset: $G^{\cal K}$ component measured at one-third of the chain and $T=0.1$. Black lines depict the $1/N$ predictions obtained by inverting the matrix in Eq.~\eqref{eq:ansatz}. The symmetry around $\omega=0$ is broken as $a$ increases. Right: Color plot of the absolute value of  $G^R(\omega=0)$  for the first $20\times20$ coarse grained cells of a system with $N=2000$ sites, darker colors represent higher values}
\end{figure*}

The first step is to fix the action of the infinite size theory $S_{\infty}$ with the aid of the coarse graining procedure. We start by disposing the elements of $G^{\mathcal{R}/\mathcal{A}/\mathcal{K}}_{i,j}$ as a matrix and subdivide it in square cells of width $a$.  We take the average over all the terms in the cell to obtain the effective Green function $G^{\mathcal{R}/\mathcal{A}/\mathcal{K}}_{\tilde i, \tilde j}$, describing the correlations between the $\tilde{i}$ and $\tilde{j}$th cell.  This procedure is illustrated in Fig.~\ref{fig:RG}-(right) for the retarded Green's function and increasing cell size ($a=1$ corresponds to no coarse graining). 
As the cell size increases, $G^{\mathcal{R}/\mathcal{A}/\mathcal{K}}_{\tilde{i},\tilde{j}}$ becomes a diagonal matrix with the off-diagonal terms vanishing as $1/a$ and exponentially suppressed with the distance $|\tilde{i}-\tilde{j}|$.  

This explicit calculation confirms the diagonal structure of $G^{\mathcal{R}/\mathcal{A}/\mathcal{K}}$ and the reduction of the action to a  sum of local commuting terms $S_{\infty}=\sum_{\tilde{j}} S_{\tilde{j}}$, where $S_{\tilde{j}}$ is the action associated to the $\tilde{j}$th cell.  To simplify the notations, we drop the tilde indices from now on and implicitly assume that the calculations are done in the effective coarse-grained theory.
The diagonal terms of $G^{\mathcal{R}}(\omega=0)$ are depicted in Fig.~\ref{fig:RG}-(left) as function of frequency with $G^{\mathcal{K}}$ shown in the inset.  As $a$ increases, the symmetry center of the functions changes to $\omega=-2\tau $  converging to the black curves depicting Eqs.~\eqref{eq:GKinfinity},\eqref{eq:GRinfinity}. As mentioned before, the only free parameters that need to be fixed in the local theory are $\mu_j, T_j$ and $\Sigma_j(\omega)$. For the dephasing model, we find that the self energy is simply given by $i\gamma_{\rm Dph}/2$. For a single site such an imaginary term was shown \cite{JinFilipponeGiamarchi_GenericMarkovian} to coincide with the effective action of a reservoir within the limit $\mu_j,T_j\to \infty$ while keeping the ratio $\mu_j/T_j$ fixed. Let $n_j$ be the local density at site $j$, $n_j =\frac{1}{2}(1-iG^K(t,t))$. Using $[G^{-1}]^\mathcal{K}=-G^{\mathcal{R}-1}G^\mathcal{K}G^{\mathcal{A}-1}$ and $G^\mathcal{K}(\omega)=-\tanh{\frac{\mu_j}{T_j}}(G^\mathcal{R}(\omega)-G^\mathcal{A}(\omega))$. Interestingly, at leading order in $1/N$, this relation turns out to be verified even at the microscopic level, i.e for $a=1$. This tells us that the local equilibration condition of the infinite size theory is always true in our case. We furthermore suppose that in the coarse-grained theory, the expression of the retarded and advanced components will be given by a single-site two-level system, i.e we suppose the following expression for $S_{\tilde{j}}$:

\begin{align}\label{eq:ansatz}
S_{\tilde{j}} & =\int\frac{d\omega}{2\pi}(\bar{\psi}_{j}^{1},\bar{\psi}_{j}^{2})\begin{pmatrix}\omega+i\frac{\gamma_{\rm Dph}}{2} & -i(2n_{j}-1)\gamma_{\rm Dph}\\
0 & \omega-i\frac{\gamma_{\rm Dph}}{2}
\end{pmatrix}\begin{pmatrix}\psi_{j}^{1}\\
\psi_{j}^{2}
\end{pmatrix}
\end{align}
Where we absorbed the $-2\tau$  shift of frequencies in the integral. Expression \eqref{eq:ansatz} is valid in the bulk, independently from any value of $\mu,T$ at the boundaries. We check explicitly that the coarse-grained theory indeed converges towards $S_{\tilde{j}}$ as $a$ is increased as shown in Fig. \ref{fig:RG}.

In the path integral formalism, the $1/N$ corrections to the current \eqref{eq:perturbativecurrent} is given by \begin{equation}\label{eq:pert1/N}
J=i\langle\hat{J}_{j}[\bar{\psi}^{+},\psi^{+}]S_{{\rm dyn}}\rangle_{\infty}
\end{equation}
where $\hat{J}$ is the current operator, $\hat{J}[\bar{\psi}^{+},\psi^{+}]$
is the evaluation of this operator in the fermionic coherent basis
on the $+$ Keldysh contour, $\langle\bullet\rangle_{\infty}:=\int{\cal D}[\psi^{\pm},\bar{\psi}^{\pm}]e^{iS_{\infty}}\bullet$ and $S_{{\rm dyn}}$ is the Keldysh action \eqref{eq:actiondyn}  associated to the contour integral of $\hat{H}_{{\rm dyn}}$ defined in \eqref{eq:perturbativecurrent}. Here we have explicitly that $\hat{H}_{{\rm dyn}}=\tau\sum_j c_j^\dagger c_{j+1} +\rm{h.c}$. The current operator is in this case : 
\begin{equation}
\hat{J}_{j}=i\tau(c_{j+1}^{\dagger}c_{j}-c_{j}^{\dagger}c_{j+1}).
\end{equation}
A straightforward calculation reported in Appendix~\ref{sec:computeationcurrent1/N} then leads to an explicit  derivation of Fick's law: 
\begin{align}\label{eq:solFick}
J & =-\frac{2\tau^{2}}{\gamma_{{\rm Dph}}}\nabla n_{j}.
\end{align}
where $\nabla$ is the discrete gradient $\nabla n_{j}=n_{j+1}-n_{j}$. 
Equation~\eqref{eq:solFick}, derived from the $1/N$ expansion, coincides with the exact result~\eqref{eq:ddeph} in the whole parameter space. 
Such agreement validates the $1/N$ expansion as a systematic and efficient procedure to compute diffusion constants. From the computational point-of-view, note that the $1/N$ expansion did not resort to any numerical schemes and provided an exact expression of the diffusive constant, which could not be extracted explicitly from the Dyson equation \eqref{eq:Dyson}.

\subsection{QSSEP} \label{sec:QSSEP}

In this section, we illustrate how our method can also be applied
to the study of the \emph{quantum symmetric simple exclusion process} (QSSEP)~\citep{BernardJin_QSSEP}. 

The QSSEP is a model of fermionic particles that hop on the lattice with random
amplitudes which can be thought as the quantum generalization of classical
exclusion processes \citep{Derrida_ReviewSSEP}. Classical exclusion
processes have attracted a widespread interest over the last decades
as they constitute statistical models with simple rules but a rich
behavior that is thought to be representative of generic properties
of non-equilibrium transport. It has been particularly impactful in
the formulation of the macroscopic fluctuation theory (MFT) \citep{MFT_Review},
which aims at understanding in a generic, thermodynamic
sense, macroscopic systems driven far from equilibrium.
It is hoped that the QSSEP will play a similar role in a quantum version of MFT, which is for now  largely unknown.

We are interested in a model of QSSEP plus the coherent jump Hamiltonian \eqref{eq:coherentjumpHamiltonian} that was first studied in Ref.~\cite{Eisler_CrossoverBallisticDiffusive}. The case of pure QSSEP can be retrieved in the limit $\tau\to0$. As for the dephasing model discussed in Sec.~\ref{sec:DPH}, we will see that the $1/N$ expansion formalism again offers a simple route to derive the diffusive current.

As pictured in Fig.~\ref{fig:differentmodels}, the QSSEP couples nearest neighbour sites. It is derived from  Eqs.~\eqref{eq:Liouvillian} and~\eqref{eq:stochHamiltonian} by taking the prescription 
\begin{equation}
    \gamma_{i,j}=\gamma_{\rm QS}\,\frac{\delta_{i,j+1}+\delta_{i,j-1}}2\,.
\end{equation}
The associated QSH is 
\begin{equation}
dH_{t}=\sqrt{\g_{\text{QS}}}\sum_{j}\left[c_{j}^{\dagger}c_{j+1}dW_{t}^{j+1,j}+c_{j+1}^{\dagger}c_{j}dW_{t}^{j,j+1}\right]\,.
\end{equation}
From Eq.~\eqref{eq:self}, we get the advanced and retarded components
of the self energies: 
\begin{equation}\label{eq:selfQSSEP}
\Sigma_{j,k}^{\cal{R}(\cal{A})}(t,t')=\mp \frac{i}{2}\gamma_{\rm QS}\delta_{j,k}\delta(t,t')\left[1-\frac{\delta_{1,j}+\delta_{j,N}}2\right]\,.
\end{equation}
The retarded and advanced Green functions are given by inserting the bare propagators \eqref{eq:bareprop} and the self energy \eqref{eq:selfQSSEP} into the Dyson equation \eqref{eq:Dyson} . These propagators can be directly derived from the ones of the dephasing model by making the substitutions $\gamma_{\rm Dph}\rightarrow \gamma_{\rm QS}$ and $\Gamma\rightarrow\Gamma-\gamma_{\rm QS}/2$. As a consequence, all the considerations made for the spectral function and Fig.~\ref{fig:spectral_dephasing}, in the dephasing model, equally apply to the QSSEP.

\begin{figure}
\begin{center}
\includegraphics[width=0.90\columnwidth]{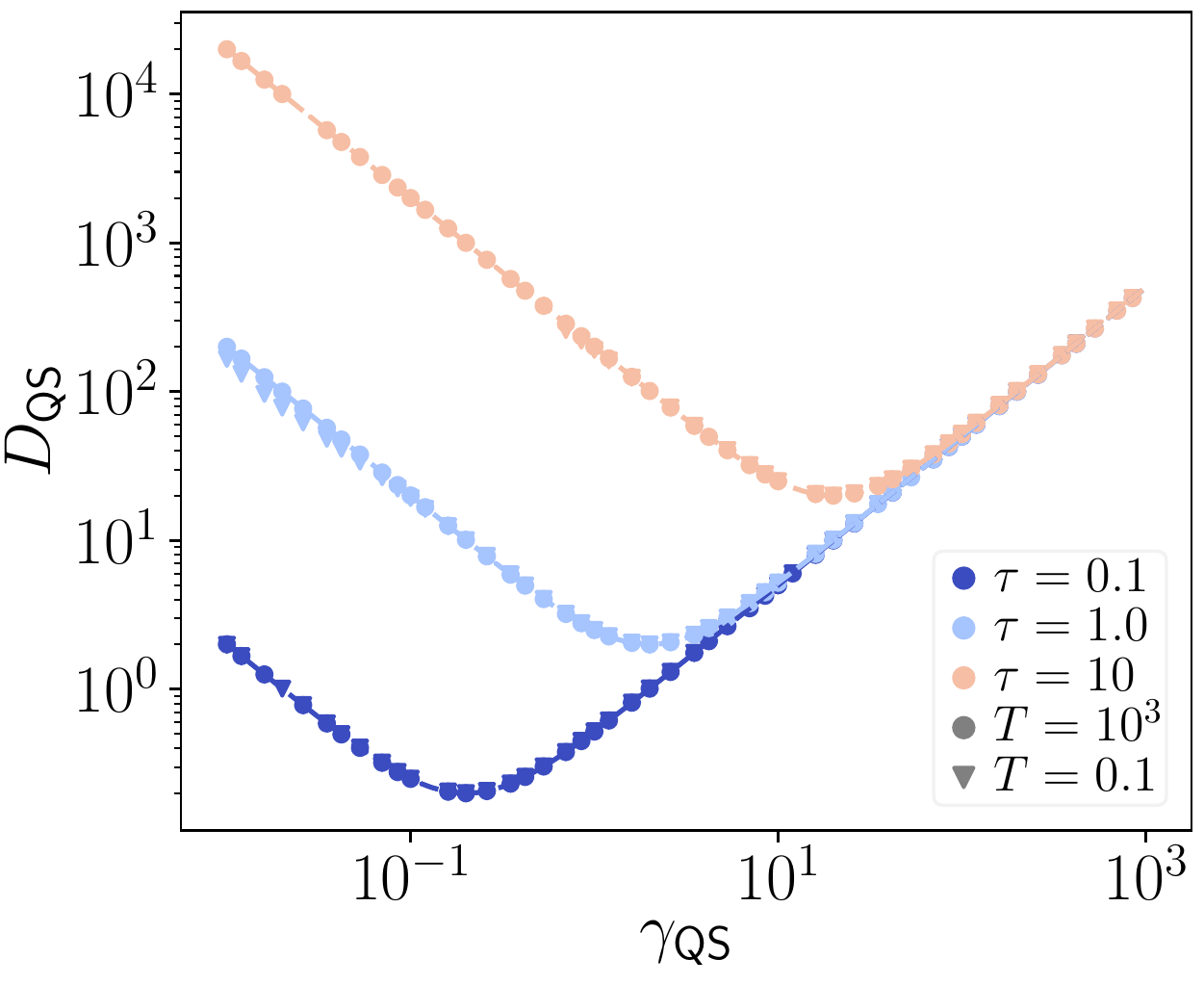}
\end{center}
\caption{\label{fig:diff_QS} 
Diffusion constant of the QSSEP model as a function of the noise strength, $\g$, for different hopping amplitudes $\tau$ and temperatures $T$. The results are independent of the latter. The dots are obtained from the MW formula \eqref{eq:MWDn} while dashed lines depict the results of the $1/N$ expansion $\eqref{eq:currentcoherentQSSEP}$.}
\end{figure}

This is not the case for the Keldysh component, where the $M$ matrix has the different expression~\footnote{in the previous expression, if an index is out of boundary, it must simply be set to $0$, we don't write that explicitly to avoid cumbersome notation.} 
\begin{align}
M_{j,k}=\frac{\gamma_{{\rm QS}}}{2}\int\frac{d\omega}{2\pi}\big(&G_{j,k-1}^{\cal R} G_{k-1,j}^{\cal A} + G_{j,k+1}^{\cal R} G_{k+1,j}^{\cal A}\big).
\end{align}
Combining the above equation with (\ref{eq:GK}) allows to obtain $G^{K}$ and allows to 
compute the current from \eqref{eq:MWDn}, or its linearized version~\eqref{eq:linearreponse}. For all values of the parameter space $(T,\gamma_{{\rm QS}})$
the current follows the relation (see Fig.~\ref{fig:diff_QS})
\begin{equation}
J_{j}=-\left(\frac{\gamma_{{\rm QS}}}{2}+\frac{2\tau^{2}}{\gamma_{{\rm QS}}}\right)\nabla n_{j}.\label{eq:currentcoherentQSSEP}
\end{equation}
which tells us that the diffusion constant is $\frac{\gamma_{{\rm QS}}}{2}+\frac{2\tau^{2}}{\gamma_{{\rm QS}}}$ in agreement with the result presented in \cite{Eisler_CrossoverBallisticDiffusive}. 
For $\tau=0$, this generalizes the result from \cite{BernardJin_QSSEP} which was restricted to boundaries with infinite temperature and chemical potential. 

\subsubsection*{1/N expansion} 
The expression (\ref{eq:currentcoherentQSSEP}) for the current can also be obtained easily in the $1/N$ perturbative approach illustrated in Sec. \ref{sec:Analy-D}. The action in the infinite size limit is again of the form \eqref{eq:ansatz}. From \eqref{eq:selfQSSEP} we see that the expression of the self-energy is similar to the one of the dephasing model by simply replacing $\gamma_{\rm Dph}$ by $\gamma_{\rm QS}$ up to differences that tend to $0$ in the infinite size limit. The current operator from site $j$ to $j+1$ in the bulk is given here by 
\begin{equation}
\hat{J}_{j}=\frac{\gamma_{{\rm QS}}}{2}(\hat{n}_{j}-\hat{n}_{j+1})+i\tau(c_{j+1}^{\dagger}c_{j}-c_{j}^{\dagger}c_{j+1}).
\end{equation}
The first part is easily evaluated to be $-\gamma_{{\rm QS}}\nabla n_{j}/2$ to first order in $1/N$ in the diffusive limit.
For the second part, we simply need to redo the previous derivation
by replacing $\gamma_{{\rm Dph}}$ by $\gamma_{{\rm QS}}$. The term $i\tau(c_{j+1}^{\dagger}c_{j}-c_{j}^{\dagger}c_{j+1})$ then becomes $-\frac{2\tau^2}{\gamma_{\rm QS}}(n_{j+1}-n_{j})$ which yields $\eqref{eq:currentcoherentQSSEP}$.

\subsection{Long-range Hopping \label{sec:LR}}

Finally we turn to the model with long-range hopping from the noise (see Fig.~\ref{fig:differentmodels}). In this model each particle can jump to any unoccupied site with a probability rate that decays with the distance as a power law of exponent $\alpha$. Power-laws appear naturally for instance in quantum simulation with Rydberg atoms \citep{Qsimulation1,Qsimulation2,Qsimulation3} where they emerge because of the dipole-dipole interactions. Depending on the order of the interactions between atoms, different power laws can be reached. In the limit $\alpha \to \infty$, we get an ``all-to-all'' model, i.e there are random quantum jumps between all sites. These types of models have recently attracted interest as toy models to understand the interplay between quantum chaos and quantum information notably in the context of random unitary circuits \citep{Nahum_RandomUnitary,Nahum_AlltoAllRandomunitary}. 

For the long-range QSH we have 
\begin{equation}
    \gamma_{i,j}=(1-\delta_{i,j})\frac{\gamma_\text{LR}}{\mathcal{N}_\alpha |j-k|^\alpha}
\end{equation}
and the corresponding Hamiltonian is
\begin{equation}\label{eq:H_LR}
dH_{t}=\sum_{j\neq  k}\sqrt{\frac{2\g_{\text{LR}}}{\mathcal{N_{\alpha}}|j-k|^{\alpha}}}c_{j}^{\dagger}c_{k}dW_{t}^{k,j}. 
\end{equation}
where $\mathcal{N_{\alpha}}=2 \sum_{k=1}^{N/2}k^{-\alpha} $ is a suitable normalization condition such that $\mathcal{N}_{\inf}=2$ and $\mathcal{N}_{0}=N$. The limiting cases of this model are the QSSEP and "all-to-all" model, respectively $\alpha=0$ and $\alpha \ra \inf$. 

For the long-range hopping the expression of the retarded(advanced) self-energy is 
\begin{equation}
    \Sigma_{j,k}^{\cal R(A)}(t,t')=\mp\delta_{j,k}\delta(t-t')\frac{i}{2}\sum_{l\neq j}\frac{\gamma_{\rm LR}}{\mathcal{N}_\alpha |j-l|^\alpha}. \label{eq:selfLRH}
\end{equation}
As before, injecting the bare propagators \eqref{eq:bareprop}, \eqref{eq:bareprop2} and \eqref{eq:selfLRH} in \eqref{eq:Dyson} yields $G^{\cal R(A)}$. As illustrated in Fig.~\ref{fig:ltLRH} in Appendix~\ref{app:longrange}, this form of the self-energy is equivalent to the one derived for the dephasing model~\eqref{eq:selfdph}, with the only difference that the effective dephasing rate $\gamma$ becomes site-dependent because of the presence of boundaries connected to reservoirs . We verified that the exponential convergence of the spectral function illustrated in Fig.~\ref{fig:spectral_dephasing}, equally applies, as expected, for this model as well.

The $M$ matrix is 
\begin{equation}
    M_{j,k}=\sum_{l \neq k} \int \frac{d\omega}{2\pi} G^{\mathcal{R}}_{j,l}(\omega)G^{\mathcal{A}}_{l,j}(\omega)\frac{\gamma_{\rm LR}}{\mathcal{N}_\alpha |k-l|^\alpha}
\end{equation}
which combined to \eqref{eq:GK} yields $G^{K}$. 

In the absence of coherent hopping, there is a simple argument to conjecture a phase transition in the transport properties of the system at $\alpha=3$. If one considers the stochastic process \eqref{eq:H_LR} alone, its average has a simple interpretation as a classical Markov process, where the probability for a fermion at site $0$ to jump to site $j$ during a timestep $\Delta t$, given that the target site $j$ is empty, is $p_j :=\frac{\gamma_{\rm LR}}{\mathcal{N}_\alpha |j|^\alpha}\Delta t$. For a single particle, this defines a random walk whose variance is given by $v:=\sum_j p_j j^2$ which is related to the diffusion constant via $D=v/\Delta t$. This diverges at least logarithmically for $\alpha \leq 3$. However, note that there is no simple reasoning to understand what happens if one were to study the model with the coherent hopping term as, {\it a priori}, a purely classical analysis does not hold anymore.

\begin{figure}
\begin{center}
\includegraphics[width=0.95\columnwidth]{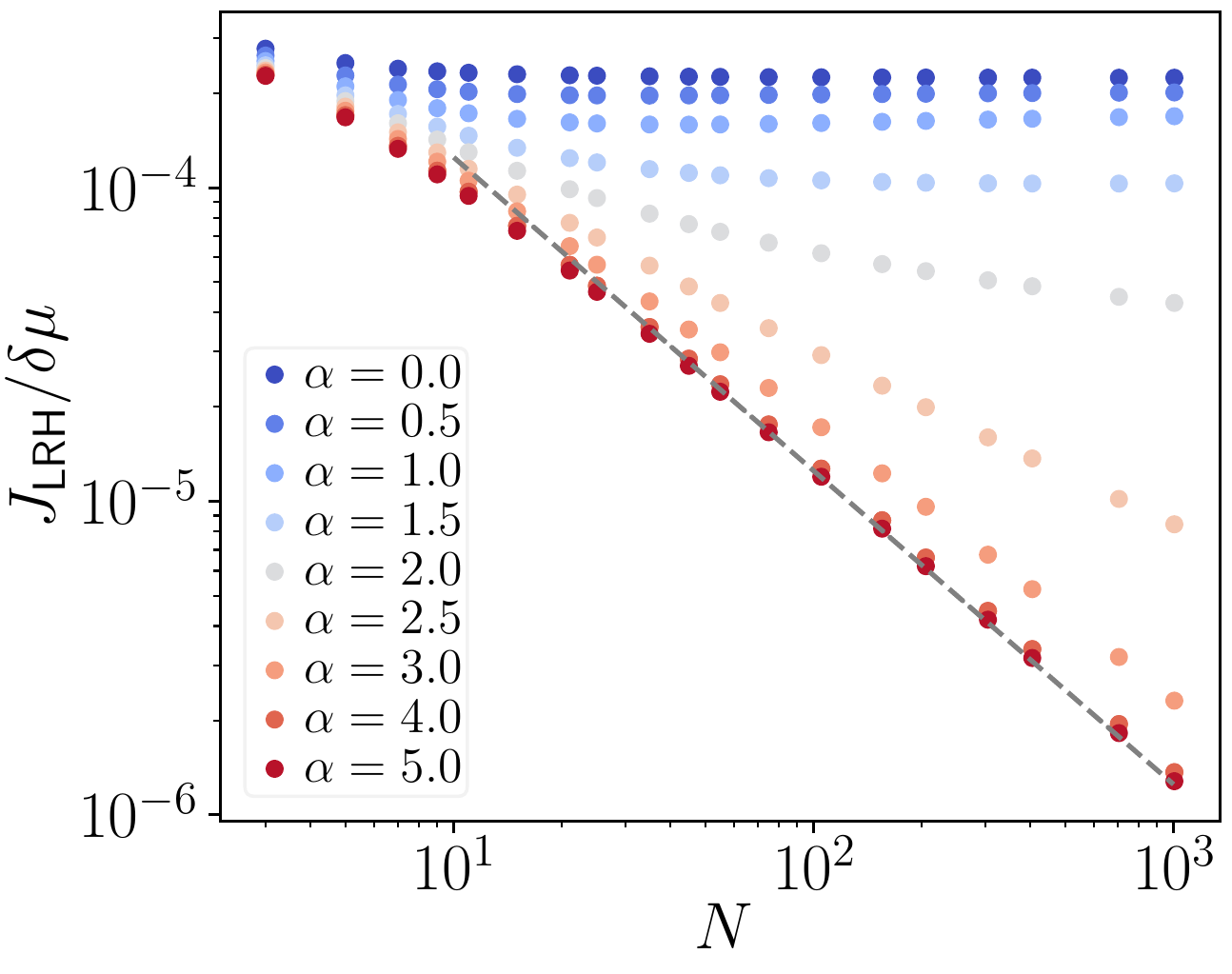}
\end{center}
\caption{\label{fig:current_L_alpha} 
Scaling of the linear response current as a function of the system size $N$, for varying power-law coefficients $\alpha$ in the long-range hopping Hamiltonian~\eqref{eq:H_LR}. The saturation of $J_{\text{LRH}}$ to finite values, for $\alpha \ll 1$, signals a ballistic regime of transport, which  contrasts with the diffusive regime observed for $\alpha \gg 1$, where $J_{\rm LRH}$ vanishes as $N^{-1}$,  as  highlighted by the black dashed line.}
\end{figure}

For the numerical computations, we fix $\g_{\text{LR}}=1$ and $T=1000$ but the results are independent of the latter. In Fig.~\ref{fig:current_L_alpha}, we show the dependence of the linear response current with the system size for different values of $\alpha$. 
When $\alpha$ is small, the current saturates in the $N\ra \inf$ limit, while for large values of $\alpha$ it decays as $N^{-1}$, as depicted in dashed gray line. This a signature of a ballistic-to-diffusive transition that occurs at a finite value of $\alpha$.

To characterize this transition further, we look at the order parameter $D^{-1}=-\lim_{N\ra \inf }\nabla n/J$. For diffusive systems, $D^{-1}$ is the inverse of the diffusion constant and should be zero for ballistic systems. In App.~\ref{sec:App-Numerics_2}, we discuss the numerical fitting required to obtain $D^{-1}$ from a finite-size scaling analysis. $D^{-1}$ undergoes a second order phase transition at a critical power $\alpha_c\approx 2.87$  (see the dark-blue dots in Fig.~\ref{fig:invD_alpha}). When approaching the transition from the diffusive region, the diffusion constant diverges as $D\sim (\alpha-\alpha_c)^{1.21}$ (see the gray dashed line in Fig.~\ref{fig:invD_alpha}).
It is quite remarkable and counterintuitive that setting $\tau \neq 0$ pushes the diffusive regime to values of $\alpha<3$ instead of the opposite. A naive reasoning would suggest that the addition of a coherent hopping term would push the  ballistic phase to values of $\alpha$ larger than the classical estimate ($\alpha=3$), as a finite $\tau$ would favor the coherent propagation of single particles across the system. We observe that the opposite is surprisingly true, and we leave the exploration of this effect to future investigations.

\subsubsection*{1/N expansion} 
For $\alpha>\alpha_c$, the $1/N$ expansion is valid and we can compute $D^{-1}$ in the limit of infinite temperature. The action in the infinite system size is again of the form \eqref{eq:ansatz} and the lifetime is fixed by \eqref{eq:selfLRH}.

\begin{figure}
\begin{center}
\includegraphics[width=0.95\columnwidth]{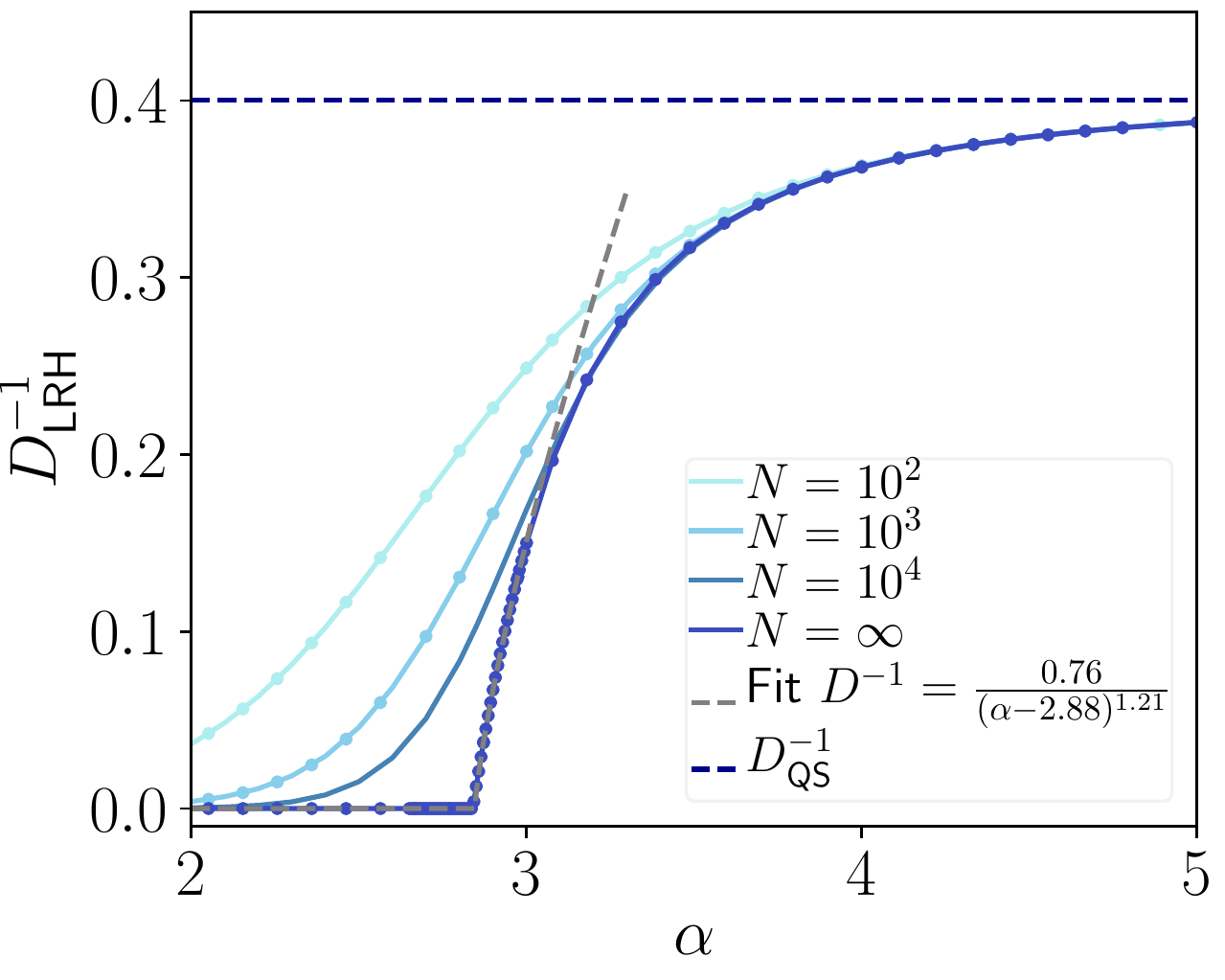}
\end{center}
\caption{\label{fig:invD_alpha} 
Second order phase transition in the long-range hopping of $D^{-1}=-\lim_{N\ra \inf }\nabla n/J_{\text{LRH}}$ as a function of $\alpha$ and $\gamma_{\text{LR}}=1$. Dots represent the numerical solution of \eqref{eq:linearreponse} while full lines depict the $1/N$ expansion's predictions; both results overlap. The $N\ra \inf$ limit is obtained via the fitting procedure detailed in Appendix~\ref{sec:App-Numerics_2}. The gray dashed line highlights the divergence of the diffusion constant as $D\sim (\alpha-\alpha_c)^{1.21}$.}
\end{figure}

Unlike the previous models, there is no simple analytic expression for the diffusion constant since its derivation depends on the system size. We provide a detailed derivation of the diffusive current in App.~\ref{sec:computeationcurrent1/N}. In Fig.~\ref{fig:invD_alpha}, we depict the results of the $1/N$ expansion for various system sizes (full lines) and overlap them with the numeric solution of \eqref{eq:linearreponse} (dots). 
Both methods agree up to machine precision which may be an indication that the $1/N$ perturbative approach is surprisingly exact even in the ballistic regime, $\alpha<\alpha_c$.

As already highlighted above, the interplay between transport and coherence  gives rise to a rich physics in the long-range hopping model, but understanding it in depth is beyond the goals of this paper and will be addressed in a subsequent work. 

\section{Conclusion}\label{sec:Conclusion}

In this work, we provided a comprehensive analysis of the large system size properties of diffusive quantum systems driven out-of-equilibrium by boundary reservoirs. In particular, we showed that diffusive quantum systems can be described by an effective and simple equilibrated Gaussian theory, which allows a systematic way to compute their diffusive transport properties via an expansion in the inverse system size. We illustrated the correctness of our $1/N$ expansion by comparing to exact results we obtained, using a self-consistent Born method, for a large class of quantum stochastic Hamiltonians which show diffusive behavior. In particular, the self-consistent approach allowed us to explicitly derive the structure of the effective Gaussian theory, which consists of decoupled sites with a finite lifetime and where the effective equilibration and diffusivity is entirely encoded in the Keldysh component of local correlations. 

As an illustration of the effectiveness of our approach, we computed the current in three models that have been of interest in the recent literature: the dephasing model, the QSSEP and a model with stochastic long-range hopping. For the dephasing model and the QSSEP, we illustrated the ability of our approach to extend the study of transport to situations with  boundaries at finite temperatures and arbitrary chemical potentials. This allowed us to show how dissipative processes restore effective infinite temperature behavior in the bulk and explicitly derive the effective Gaussian theory via a coarse-graining procedure. For the long-range hopping model, our analysis unveiled that coherent hopping processes trigger diffusive behavior in regimes where transport would be ballistic in the exclusive presence of stochastic long-range hopping. This counter-intuitive phenomenon is a remarkable example of the non-trivial interplay between coherent and dissipative dynamics in open quantum systems, which could be efficiently addressed based on the self-consistent approach.

The validity of the self-consistent Born approximation for our class of stochastic Hamiltonians provides in principle the solution to the noisy version of any model whose bare action is Gaussian. 
Our proof  is not limited by stationary behavior or by the  one-dimensional geometry of the problems addressed in this paper, but can be extended to time-dependent and higher dimensional problems as well. This possibility opens interesting perspectives for the investigation of novel phenomena in a large class of problems. Extension of our approach could be devised to study quantum asymmetric exclusion processes~\citep{HudsonParthasarathy_QuantumIto,Derrida_SolutionASEP,JinKrajenbrinkBernard_QKPZ}, spin and heat transport, the dynamics after a quench, fluctuations on top and relaxation to stationary states and their extensions to ladder geometries or with non-trivial topological structure.  These settings have been for the moment largely untractable, or were solved by case by case methods, for which we provided here an unified framework.  

An important issue raised by our work consists in  showing whether our description equally holds and provides technical advantage for studying the emergence of resistive behavior triggered by intrinsic many-body interactions with unitary dynamics, where  the breaking of integrability leads to diffusive transport~\cite{giamarchi_umklapp_1d,rosch_conservation_1d,zotos_conductivity_integrable,bertinifinite2021,denardisDiffusionGeneralizedHydrodynamics2019,friedmanDiffusiveHydrodynamicsIntegrability2020,znidaricNonequilibriumSteadystateKubo2019a,znidaricWeakIntegrabilityBreaking2020a,znidaricAbsenceSuperdiffusionQuasiperiodic2020,ferreiraBallistictodiffusiveTransitionSpin2020a}.
A priori, the arguments presented in Section~\ref{sec:Analy-D} apply for any quantum
systems which follows a local Fick's law and, as such, they have the potential
for very broad applications. 
Additionally, it is commonly accepted that the phenomenology of diffusion is associated with integrability breaking and subsequent approach to thermal equilibrium \cite{deutschQuantumStatisticalMechanics1991,srednickiApproachThermalEquilibrium1999,rigolThermalizationItsMechanism2008,dalessioQuantumChaosEigenstate2016,kinoshitaQuantumNewtonCradle2006}. Understanding if and how our approach can help make this link clearer is an exciting open question. In this respect, we also note  that, because of the existing mapping between the Fermi-Hubbard
and the dephasing model \citep{ProsenEssler_Mapping}, the self-consistent Born approximation allows to compute exact quantities in the Fermi-Hubbard model. As far as we know, exact solutions for this model were only obtained
in the framework of the Bethe Ansatz and it is thus interesting that
a seemingly unrelated approach allows to obtain exact quantities as
well. Whether a connection exists between the two approaches and whether the exact summation allows to compute quantities out of reach of the Bethe ansatz are interesting open questions.

\begin{acknowledgments}
We thank L. Mazza for useful suggestions during the writing of the manuscript. This work has been supported by the Swiss National Science Foundation under Division II. J.F. and M.F.  acknowledge  support  from  the  FNS/SNF  Ambizione  Grant PZ00P2\_174038. 

We also thank  X.  Turkeshi and M. Schir\'o  for making us aware, in the final phase of the writing of this manuscript, of their work~\citep{TurkeshiSchiro_Dephmodel} before publication, where a study of the dephasing model from the point of view of Green's function has also been performed.
\end{acknowledgments}

\appendix

\section{Unraveling to continuous measurement}\label{app:measurement}

In this appendix, we discuss the unraveling of Eq.~\eqref{eq:Liouvillian} to a quantum stochastic differential equation describing a system under continuous monitoring. In the It\={o} prescription the stochastic equation of motion of a quantum system subject to continuous measurement
of an observable $O+O^{\dagger}$ at rate $\gamma$ is given by \citep{Belavkin_Nondemolitionmeasurements}
\begin{equation}
d\rho={\cal L}_{0}(\rho)+\frac{\gamma}{2}L_{O}(\rho)+\sqrt{\frac{\gamma}{2}}D_{O}(\rho)dB_{t}\label{eq:measurement}
\end{equation}
where $\mathcal{L}_0$ describes the dynamics in absence of measurement, $L_{O}(\rho)=(O\rho O^{\dagger}-\frac{1}{2}(O^{\dagger}O\rho+\rho O^{\dagger}O))$
and $D_{O}(\rho)=O\rho+\rho O^{\dagger}-\rho{\rm tr}(O\rho+\rho O^{\dagger})$.
If we assume that at each link we have two independent measurement
processes 1 and 2 with the same rate $2\gamma_{i,j}$ and $O_{1,i,j}:=c_{j}^{\dagger}c_{i}$
and $O_{2,i,j}:=ic_{j}^{\dagger}c_{i}$. The corresponding measured
observables are $O_{1,i,j}+O_{1,i,j}^{\dagger}=c_{j}^{\dagger}c_{i}+c_{i}^{\dagger}c_{j}$
and $O_{2,i,j}+O_{2,i,j}^{\dagger}=i(c_{j}^{\dagger}c_{i}-c_{i}^{\dagger}c_{j})$, 
namely the so-called bond density and the current. 
It is straightforward to see that averaging out (\ref{eq:measurement}), we get (\ref{eq:Liouvillian})
again.

\section{Proof of the non-crossing rule}\label{sec:App-Stochastic}

We want to prove that for all stochastic Hamiltonians of the form given by (\ref{eq:stochHamiltonian}),
the only non vanishing diagrams in the averaged perturbative expansion
of the retarded, advanced and Keldysh Green functions are those for
which there is no crossing.

This statement only relies on the causality structure of the retarded
and advanced Green functions, i.e 
\begin{align}
G^{\cal R}(t,t') & = 0\text{\text{ if }t\ensuremath{< t'},}\\
G^{\cal A}(t,t') & = 0\text{ if t\ensuremath{> t'}}.
\end{align}
Let $\langle\bullet\rangle_{0}$ denote the average with respect to
a quadratic theory. First, we remark that the causality structure
of a given propagator depends only on its incoming edge and outgoing edge, and thus
\begin{align}
G(t,t') & :=\langle\psi^{1}(t)f[\psi^{1},\bar{\psi}^{1},\psi^{2},\bar{\psi}^{2}]\bar{\psi}^{1}(t')\rangle_{0}=0\text{ for \ensuremath{t< t'}},\\
G'(t,t') & :=\langle\psi^{2}(t)f[\psi^{1},\bar{\psi}^{1},\psi^{2},\bar{\psi}^{2}]\bar{\psi}^{2}(t')\rangle_{0}=0\text{ for \ensuremath{t> t'}.}
\end{align}
where $f[\psi^{1},\bar{\psi}^{1},\psi^{2},\bar{\psi}^{2}]$ is an
arbitrary polynomial in the Grassman variables coming from the expansion
of the stochastic action. This is straightforward to show starting from the action (\ref{eq:stochasticaction}):
starting from an incoming full (dashed) line, one cannot switch
at any point to a dashed (full) line. Hence, the causality structure
is preserved for each line and thus for the whole propagator. Direct inspection of these diagrams show that there cannot be any crossing when contracting the noise terms, as it would lead to a contradiction
in the time-orderings. There is only a single one particle irreducible diagram made of a single loop. 
This establishes the non-crossing result for the retarded and advanced components.

\begin{figure}
\begin{center}
\includegraphics[width=0.9\columnwidth]{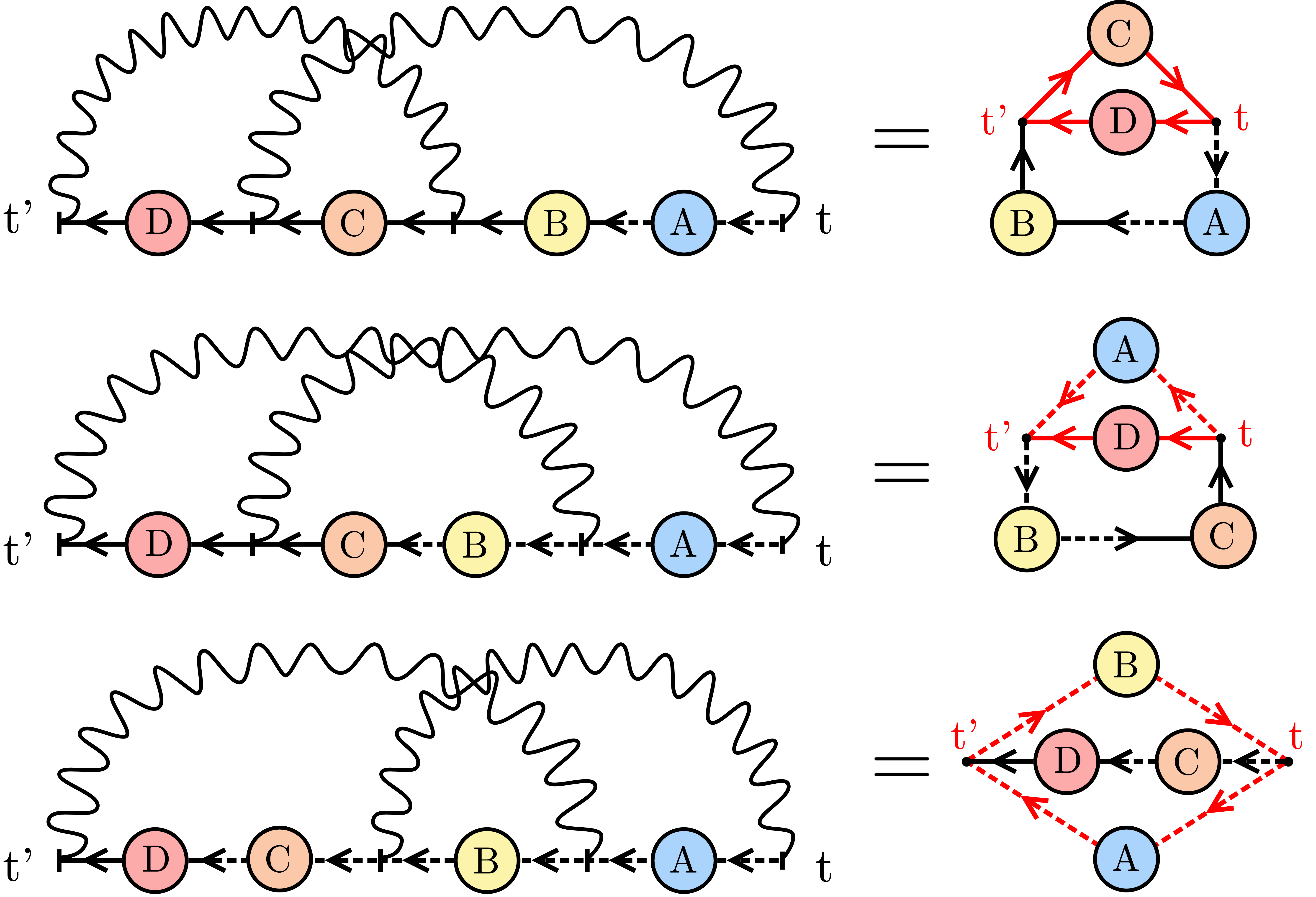}
\par\end{center}
\caption{\label{fig:crossing}
All possible crossings for the Keldysh component of the Green's function with the contracted versions on the right. The red lines highlight the part
of the diagram violating the causality structure and are responsible for making the diagram vanish.}
\end{figure}

For the Keldysh components, a case by case examination
of all possible crossings that are depicted on Fig.~\ref{fig:crossing} where the labels $A,B,C,D$
denote generic product of free propagators is needed. 
For each one of these diagrams, there is always a subpart
that shows an incompatibility (shown in red in Fig.~\ref{fig:crossing}) in the time orderings causing the whole diagram to vanish. This establishes the non-crossing result for the Keldysh propagator. 

\section{\label{sec:computeationcurrent1/N}Computation of the current in the 1/N expansion}

In this appendix, we compute the current in the dephasing, QSSEP and long-range model using the perturbative theory in inverse system size presented in Sec.~\ref{sec:Analy-D}. 

\subsection{Dephasing model}
For the dephasing model, the definition of the current in the bulk
from site $j$ to $j+1$ is given by 
\begin{equation}
\hat{J}_{j}=i\tau(c_{j+1}^{\dagger}c_{j}-c_{j}^{\dagger}c_{j+1}).
\end{equation}
The expectation value of $\hat{J}_{j}$ in the stationary state is
given by 
\begin{equation}
\begin{split}
J_{j}(t):= & {\rm tr}(\hat{J}_{j}\rho_{t}) \\
= & i\tau\langle\bar{\psi}_{j+1}^{+}(t)\psi_{j}^{+}(t)-\bar{\psi}_{j}^{+}(t)\psi_{j+1}^{+}(t)\rangle\\
= & i\frac{\tau}{2}\langle \big(\psi_{j+1}^{1}\bar{\psi}_{j}^{1}+\psi_{j+1}^{1}\bar{\psi}_{j}^{2}+\psi_{j+1}^{2}\bar{\psi}_{j}^{2} \\
 & -(\psi_{j}^{1}\bar{\psi}_{j+1}^{1}+\psi_{j}^{1}\bar{\psi}_{j+1}^{2}+\psi_{j}^{2}\bar{\psi}_{j+1}^{2})\big)_{t}\rangle
\end{split}
\end{equation}
where we used the Larkin rotation and removed the terms $\psi^{2}\bar{\psi}^{1}$
as they are always $0$ for causality reasons. 

Using the action associated to the coherent jump $S_{\tau}$
\begin{equation}
S_{\tau}=-\tau\int dt'\sum_j \big(\bar{\psi}_{j}^{1}\psi_{j+1}^{1}+\bar{\psi}_{j}^{2}\psi_{j+1}^{2}+\bar{\psi}_{j+1}^{1}\psi_{j}^{1}+\bar{\psi}_{j+1}^{2}\psi_{j}^{2}\big)_{t'}
\end{equation}
we get, from (\ref{eq:pert1/N}), to leading order in $\frac{1}{N}$: 
\begin{multline}
J_{j}(t) =  \frac{\tau^{2}}{2}\int dt'\langle\big(\psi_{j+1}^{1}\bar{\psi}_{j}^{1}+\psi_{j+1}^{1}\bar{\psi}_{j}^{2}+\psi_{j+1}^{2}\bar{\psi}_{j}^{2} \\
  -(\psi_{j}^{1}\bar{\psi}_{j+1}^{1}+\psi_{j}^{1}\bar{\psi}_{j+1}^{2}+\psi_{j}^{2}\bar{\psi}_{j+1}^{2})\big)_{t}\\
  \big(\bar{\psi}_{j}^{1}\psi_{j+1}^{1}+\bar{\psi}_{j}^{2}\psi_{j+1}^{2}+\bar{\psi}_{j+1}^{1}\psi_{j}^{1}+\bar{\psi}_{j+1}^{2}\psi_{j}^{2}\big)_{t'}\rangle_{\infty} 
\end{multline}
where $\langle\rangle_{\infty}$ means the average with respect to the
bare action in the infinite size limit, where all the sites are uncorrelated. 

Using Wick's theorem and that $\langle\psi_{j}^{a}\bar{\psi}_{j+1}^{b}\rangle_{\infty}=0$,
the previous equation greatly simplifies : 
\begin{multline}
J_{j} =  -\frac{\tau^{2}}{2}\int\frac{d\omega}{2\pi}\big(G_{j+1,j+1}^{\cal R}(\omega)G_{j,j}^{\cal K}(\omega)+G_{j,j}^{\cal A}(\omega)G_{j+1,j+1}^{\cal K}(\omega)\\
  -G_{j,j}^{\cal R}(\omega)G_{j+1,j+1}^{\cal K}(\omega)-G_{j+1,j+1}^{\cal A}(\omega)G_{j,j}^{\cal K}(\omega)\big)
\end{multline}
We can now use the  bare action of individual sites (in presence of the dephasing noise): 
\begin{equation}
S_{j} = \int\frac{d\omega}{2\pi}(\bar{\psi}_{j}^{1},\bar{\psi}_{j}^{2})\begin{pmatrix}\omega+i\frac{\gamma_{{\rm Dph}}}{2} & -i(2n_{j}-1)\gamma_{{\rm Dph}}\\
0 & \omega-i\frac{\gamma_{{\rm Dph}}}{2}
\end{pmatrix}\begin{pmatrix}\psi_{j}^{1}\\
\psi_{j}^{2}
\end{pmatrix}
\end{equation}
to obtain the explicit expression of the current
\begin{equation}
 \begin{split}
J_{j} & = \tau^{2}\int\frac{d\omega}{2\pi}\bigg(\frac{\gamma_{{\rm Dph}}}{(\omega^{2}+(\frac{\gamma_{{\rm Dph}}}{2})^{2})^{2}}\bigg)\frac{i\gamma_{{\rm Dph}}}{2}(2i(n_{j+1}-n_{j}))\\
 & = - \frac{2\tau^{2}}{\gamma_{{\rm Dph}}}\nabla n_{j}
\end{split}
\end{equation}
from which we immediately read the diffusion constant $D=\frac{2\tau^2}{\gamma_{\rm Dph}}$.

\subsection{QSSEP}
For the QSSEP, the self-energy for an individual site is $\Sigma_{j}(\omega)=\gamma_{{\rm QS}}-\frac{\gamma_{{\rm QS}}}{2}(\delta_{j,1}+\delta_{j,N})$.
The current in the bulk is given by 
\begin{equation}
 \hat{J}_{j}=\frac{\gamma_{{\rm QS}}}{2}(\hat{n}_{j}-\hat{n}_{j+1})+i\tau(c_{j}^{\dagger}c_{j+1}-c_{j+1}^{\dagger}c_{j})
\end{equation}
The first part of the current already scales like $1/N$ at order
$0$ in the $S_{\tau}$ expansion. The second term is evaluated in
the same fashion as for the dephasing model. This leads to 
\begin{equation}
J_{j}=-\left(\frac{\gamma_{{\rm QS}}}{2}+\frac{2\tau^{2}}{\gamma_{{\rm QS}}}\right)\nabla n_{j}+O\left(\frac{1}{N^{2}}\right)
\end{equation}
and $D=\frac{\gamma_{{\rm QS}}}{2}+\frac{2\tau^{2}}{\gamma_{{\rm QS}}}$.

\subsection{Long-range hopping}\label{app:longrange}
\begin{figure}
\begin{center}
 \includegraphics[width=0.80\columnwidth]{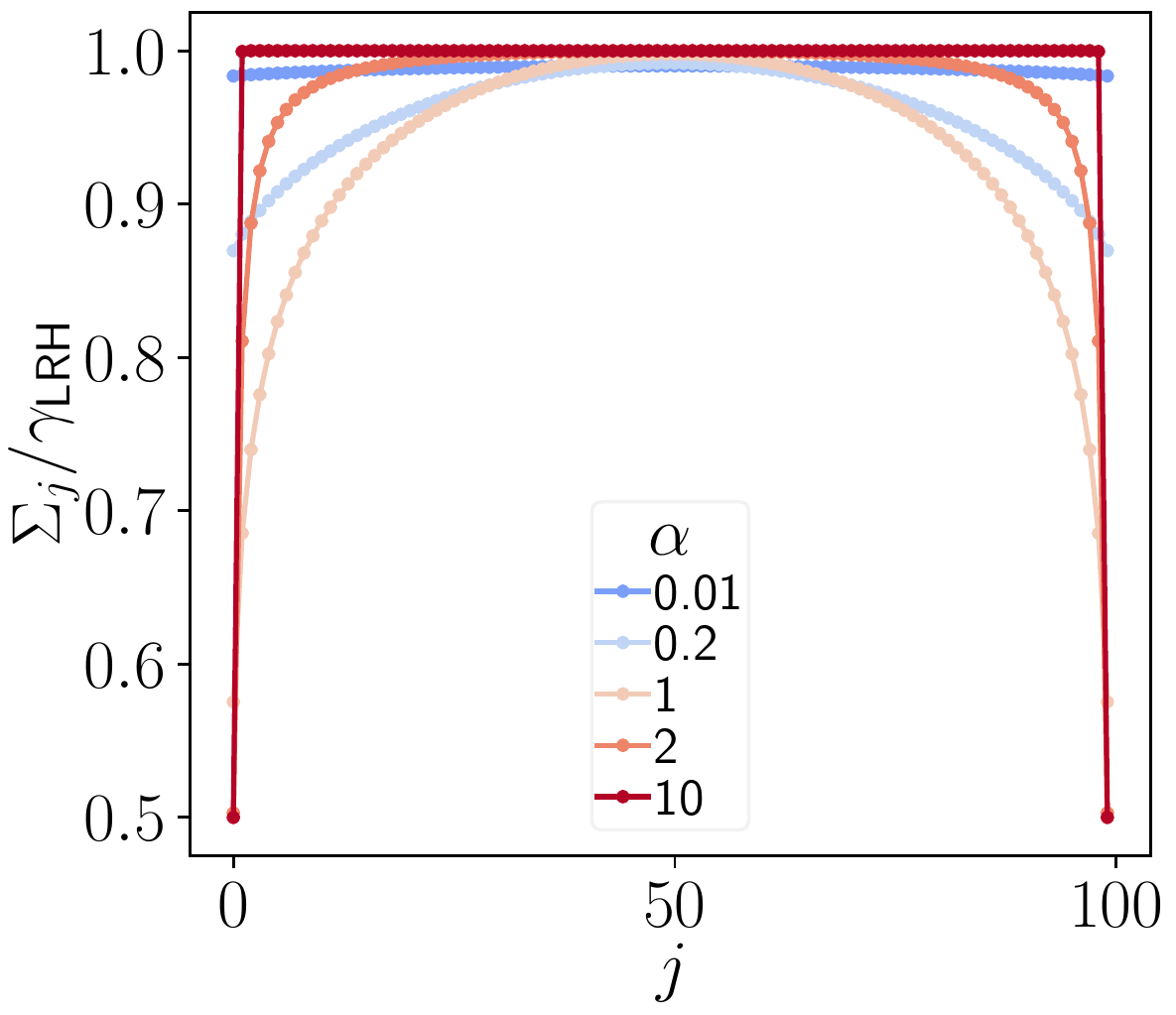}
\end{center}
\caption{\label{fig:ltLRH}
Dependence on the site index $j$ of the self-energy in the long-range model for a chain of $N=100$ sites and different values of the exponent $\alpha$ of the noise, see ~\eqref{eq:selfLRH2}. The $\alpha \ra \infty$ limit corresponds to the QSSEP.}
\end{figure}

For the long-range hopping model, the local current is defined from
the local conservation equation of the particle number : 
\begin{equation}
\frac{d}{dt}\hat{n}_{j}:=\hat{J}_{j}^{{\rm inc}}-\hat{J}_{j}^{{\rm out}}
\end{equation}
with 
\begin{align}
\hat{J}_{j}^{{\rm inc}} & =\sum_{k<j}\frac{{\cal \gamma_{{\rm LR}}}}{{\cal N}_{\alpha}|k-j|^{\alpha}}(\hat{n}_{k}-\hat{n}_{j})+i\tau(c_{j-1}^{\dagger}c_{j}-c_{j}^{\dagger}c_{j-1}),\\
\hat{J}_{j}^{{\rm out}} & =\sum_{k>j}\frac{{\cal \gamma_{{\rm LR}}}}{{\cal N}_{\alpha}|k-j|^{\alpha}}(\hat{n}_{j}-\hat{n}_{k})+i\tau(c_{j}^{\dagger}c_{j+1}-c_{j+1}^{\dagger}c_{j}).
\end{align}
Recall the expression of the self-energy at site $j$ \eqref{eq:selfLRH} :

\begin{equation}\label{eq:selfLRH2}
\Sigma_{j}=\frac{\gamma_{{\rm LR}}}{2{\cal N}_{\alpha}}\sum_{k\neq j}\frac{1}{|k-j|^{\alpha}}. 
\end{equation}
which is depicted in Fig.~\ref{fig:ltLRH}. 

To get the current with the $1/N$ expansion, we take, as for the previous model, the
$0^{{\rm th}}$ order term in the first term in the expression of
the current and the first order term in the second part. We obtain:
\begin{align}
J_{j}^{{\rm inc}}= & \sum_{k<j}\frac{\gamma_{{\rm LR}}}{{\cal N}_{\alpha}|k-j|^{\alpha}}(n_{k}-n_{j})\nonumber \\
 & +\frac{2\tau^{2}}{\Sigma_{j-1}+\Sigma_{j}}(n_{j-1}-n_{j})\text{ for }j\in[2,N],\\
\hat{J}_{j}^{{\rm out}}= & \sum_{k>j}\frac{{\cal \gamma_{{\rm LR}}}}{{\cal N}_{\alpha}|k-j|^{\alpha}}(n_{j}-n_{k})\nonumber \\
 & +i\tau(n_j-n_{j+1})\text{ for }j\in[1,N-1].
\end{align}
For simplicity, we give in this paper only the expressions for the infinite temperature and chemical potential boundary conditions which amount to take Lindblad injecting and extracting
terms (see \cite{JinFilipponeGiamarchi_GenericMarkovian}). 
The current at the boundaries is then given by: 
\begin{align}
J_{1}^{{\rm in}} & =\alpha_{L}(1-n_{1})-\beta_{L}n_{1},\\
J_{N}^{{\rm out}} & =-\alpha_{R}(1-n_{N})+\beta_{R}n_{N}.
\end{align}
In the stationary state we have that $\forall j\in[1,N]$, $J_{j}^{{\rm in}}=J_{j}^{{\rm out}}$
which leads to the following system of linear equation to solve in
order to get the density profile :
\begin{equation}
{\cal M}.\vec{n}=\vec{v}
\end{equation}
where $\vec{n}$ and $\vec{v}$ are $N$-dimensional vectors with elements
$n_{j}$ and ${\cal M}$ is an $N\times N$ matrix such that
\begin{equation}
\begin{split}
{\cal M}_{j,k} = & \frac{\gamma_{{\rm LR}}}{{\cal N}_{\alpha}|k-j|^{\alpha}}(1-\delta_{j,k})\\
 & +\frac{2\tau^{2}}{\Sigma_{j}+\Sigma_{j+1}}(\delta_{k,j+1}-\delta_{j,k}(1-\delta_{j,N})) \\
 & +\frac{2\tau^{2}}{\Sigma_{j}+\Sigma_{j-1}}(\delta_{k,j-1}-\delta_{j,k}(1-\delta_{j,1})) \\
 & -\delta_{j,k}(\sum_{k\neq j}\frac{\gamma_{{\rm LR}}}{{\cal N}_{\alpha}|k-j|^{\alpha}}) \\
 & -\delta_{j,k}\delta_{j,1}(\alpha_{L}+\beta_{L})-\delta_{j,k}\delta_{j,N}(\alpha_{R}+\beta_{R})
\end{split}
\end{equation}
and
\begin{equation}
   v_{j} =  -\delta_{j,1}\alpha_{L}-\delta_{j,N}\alpha_{R}
\end{equation}
\begin{figure*}[!htb]
\begin{center}
\includegraphics[width=0.90\textwidth]{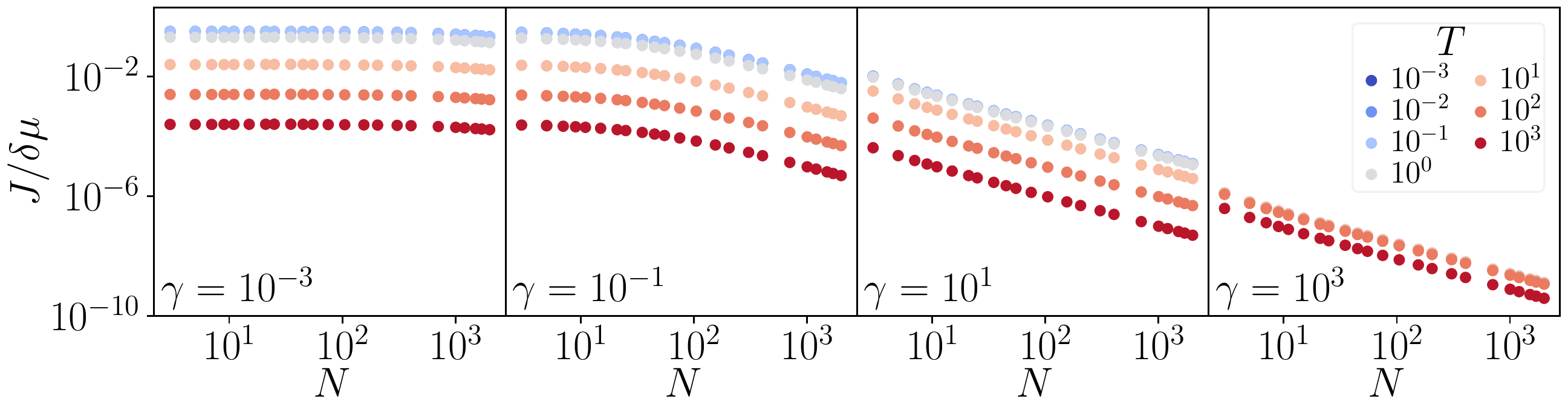}
\end{center}
\caption{\label{fig:JL_span} 
Scaling of the current as a function of the system size in the dephasing model. From left to right: $\gamma=10^{-3},10^{-1},10^{1},10^{3}$. As the dephasing increases, diffusion sets in at smaller system sizes. The vanishing dependence of $J$ with the temperature indicates the crossover into the $R_{\gamma}$ region~\eqref{eq:JTdeph}.}
\end{figure*}
\section{Numerical implementation} \label{sec:App-Numerics}

In this appendix, we present some important elements of the numerical
implementation. The first step to compute any presented result is
to stabilize and efficiently evaluate $G^{\cal{R}(\cal{A})}(\w)$ at any $\w$. For
the case of a uniform stochastic noise (e.g. free system, dephasing), a naive use of 
\eqref{eq:GRdeph} would require evaluating the ratio
of two polynomials of order $\mathcal{O}(N)$, a notoriously
difficult task for large $N$ using floating point arithmetics. A possible solution would be to
resort to arbitrary-precision arithmetic but this would entail a heavy speed cost. 

We used for the results of the present paper the fact that $G^{\cal{R}(\cal{A})}(\w)$
can be written as a ratio of polynomials and therefore, decomposed
into a product of monomials $G^{\cal R}(\w)\sim\prod_{j}(\w-z_{j})/\prod_{i}(\w-p_{i})$.
To efficiently find the zeros and poles of $G^{\cal R}$ \footnote{The poles and zeros of $G^{\cal A}$ are the conjugate of $G^{\cal R}$},
we note that the inverse of $G^{\cal R}$ is a simple tridiagonal matrix
with a generic form
\begin{equation}
T^{-1}(\w) = \left(\begin{array}{cccc}
\w+a_{1} & b_{1} & 0 & 0\\
b_{1}^{*} & \w+a_{2} & \ddots & 0\\
0 & \ddots & \ddots & b_{N-1}\\
0 & 0 & b_{N-1}^{*} & \w+a_{N}
\end{array}\right)
\end{equation}
whose inverse is given by~\cite{usmani1994inversion}
\begin{equation}
T_{i,j}(\w)=\begin{cases}
(-1)^{i+j}b_{i}...b_{j-1}\theta_{i-1}\phi_{j+1}/\theta_{L} & i<j\\
\theta_{i-1}\phi_{j+1}/\theta_{L} & i=j\\
(-1)^{i+j}b_{j}^{*}...b_{i-1}^{*}\theta_{j-1}\phi_{i+1}/\theta_{L} & i>j
\end{cases}
\end{equation}
where $\theta_{i}=(\w+a_{i})\theta_{i-1}-\left|b_{i-1}\right|^{2}\theta_{i-2}$
and $\phi_{i}=(\w+a_{i})\phi_{i+1}-\left|b_{i}\right|^{2}\phi_{i+2}$.
Therefore, computing the poles and zeros of $G^{R}$ requires computing
all the zeros of the sequences $\left\{ \phi_{i},\theta_{i}\right\} _{i=0}^{L+1}$,
a task that can be done efficiently. If the matrix is invariant under
a reflection along the anti-diagonal, it is enough to compute a single
sequence instead, $\phi_{i}=\theta_{L+1-i}$. This is always the case
in the models studied in the present paper. Since $a_{i}$ does not depend on $\w$,
$\phi_{i}$ is a polynomial of degree $i$ with the initial conditions
defined as $\phi_{0}=1$ and $\phi_{1}=\w+a_{1}$. One can efficiently
find all the roots $\{z_{k}\}_{k=1}^{i}$ of $\phi_{i}$ using a Weierstrass-like
recursive method~\cite{gargantini1971circular,petkovic2005derivative}, see Eqs. \ref{eq:WS2} and \ref{eq:WS4} for a second
and fourth order scheme
\begin{align}
z_{k}^{(2)} & =z_{k}-\frac{W_{k}}{\prod_{k\neq j}(z_{k}-z_{j})}=z_{k}-C_{k}^{(2)}\label{eq:WS2}\\
z_{k}^{(4)} & =z_{k}-\frac{W_{k}}{1-\sum_{k\neq j}\frac{W_{j}}{z_{k}-W_{k}-z_{j}}}=z_{k}-C_{k}^{(4)}\label{eq:WS4}\\
W_{k} & =\phi_{i}(z_{k})\nonumber 
\end{align}
where $W_{k}$ is the Weierstrass weight. We chose these derivative-free schemes to avoid computing explicit derivatives that would slow down the computation. Choosing the correct initial condition is critical
to the success of the scheme. To find the roots of $\phi_{i}$, we
initialize the scheme with the roots of $\phi_{i-1}$ plus an extra
root. We empirically found that the extra root should have a random
position close to the middle root (after sorting by the real part)
to guarantee the best convergence. This initial choice can still fail
when some roots are located very far way from the others, which occurs for example 
for the model QSSEP.
This happens when, at some step in the iteration, two roots coalesce
and $C_{k}^{(i)}$ diverges strongly. In order to stablize this divergence,
we introduce a damping factor $\kappa$ that suppresses large corrections $z_{i}^{(k)}=z_{i}-C_{i}^{(k)}e^{-\max|C_{i}^{(k)}|/\kappa}$. 
$\kappa$ is a purely empirically value, which we typically take as
$\kappa=\max(|b|)$. The role of $\kappa$ is to slow down the algorithm
and allow the coalescing roots to separate. Our root-searching algorithm
has thus two parts: a quick search using a second-order damped scheme,
followed by a fourth-order damped scheme to precisely locate the roots.
Once all the roots are recovered, we generate the new matrix $\tilde{T}$
obtained from the estimates of the roots. We consider that $\tilde{T}$
is a good estimate only when $\max\left|T^{-1}(0)\cdot\tilde{T}^{-1}(0)\right|<10^{-10}$.
With the exception of the QSSEP, we find a typical value $\max\left|T^{-1}(0)\cdot\tilde{T}^{-1}(0)\right|\sim10^{-13}$
for any system size.

Once the poles and zeros of $G^{\cal{R}(\cal{A})}(\w)$ are computed, we proceed
to compute $G^{\cal K}$ using \eqref{eq:GK}. To evaluate the $M$ matrix,
we resort to the residue theorem. If the poles of $G^{\cal{R}(\cal{A})}$ are simple
poles, the sum over residues can be computed in parallel only requiring
the evaluation of the monomials $\left\{ (\w-z_{k})\right\} $. We
note that while each monomial $(\w-z_{k})$ is of order unity, a sequential
multiplication can lead to overflown errors in the limit of large
$N$. To avoid this problem, we multiply the monomials at random.
If the algorithm fails to, within machine precision, separate two
roots, the residue is computed from the contour integral instead.

The last step to compute $G^{\cal K}$ and the current $J$, is to perform
the frequency integral convoluted with $\cosh^{-2}(\frac{\omega-\mu}{2T})$.
This is done by evaluating the integral using a discrete integration
scheme instead of residue theorem. Since the thermal dependence is
only encoded in the $\cosh^{-2}(\frac{\omega-\mu}{2T})$, discretizing
the integral allows us deal with different $(T,\mu)$ values at
no significant cost. We carefully verify that the mesh is fine enough
to guarantee convergence of the integral at any $(T,\mu)$.

\section{Finite-size scaling} \label{sec:App-Numerics_2}

In this section, we detail the 
finite-size scaling analysis necessary to plot Figs.~\ref{fig:current_phase_diagram-1},\ref{fig:diff_QS} and \ref{fig:invD_alpha}. 

The presence of a dephasing term is not enough to ensure that the system behaves diffusely at any system size. Signatures of diffusive transport such as $J\sim 1/N$, only emerge at a characteristic  dephasing length, $N^*\sim 1/\gamma$. At short system sizes, or short time-scales, the system behaves as if it was ballistic. In Fig.~\ref{fig:JL_span} we highlight this ballistic-to-diffusive transition for different values of the dephasing and temperature in the baths.
At small dephasing values, one cannot reliably extract the diffusion constant by fitting a a straight line to Fig.~\ref{fig:JL_span}. 
\begin{figure}[!htb]
\begin{center}
\includegraphics[width=0.95\columnwidth]{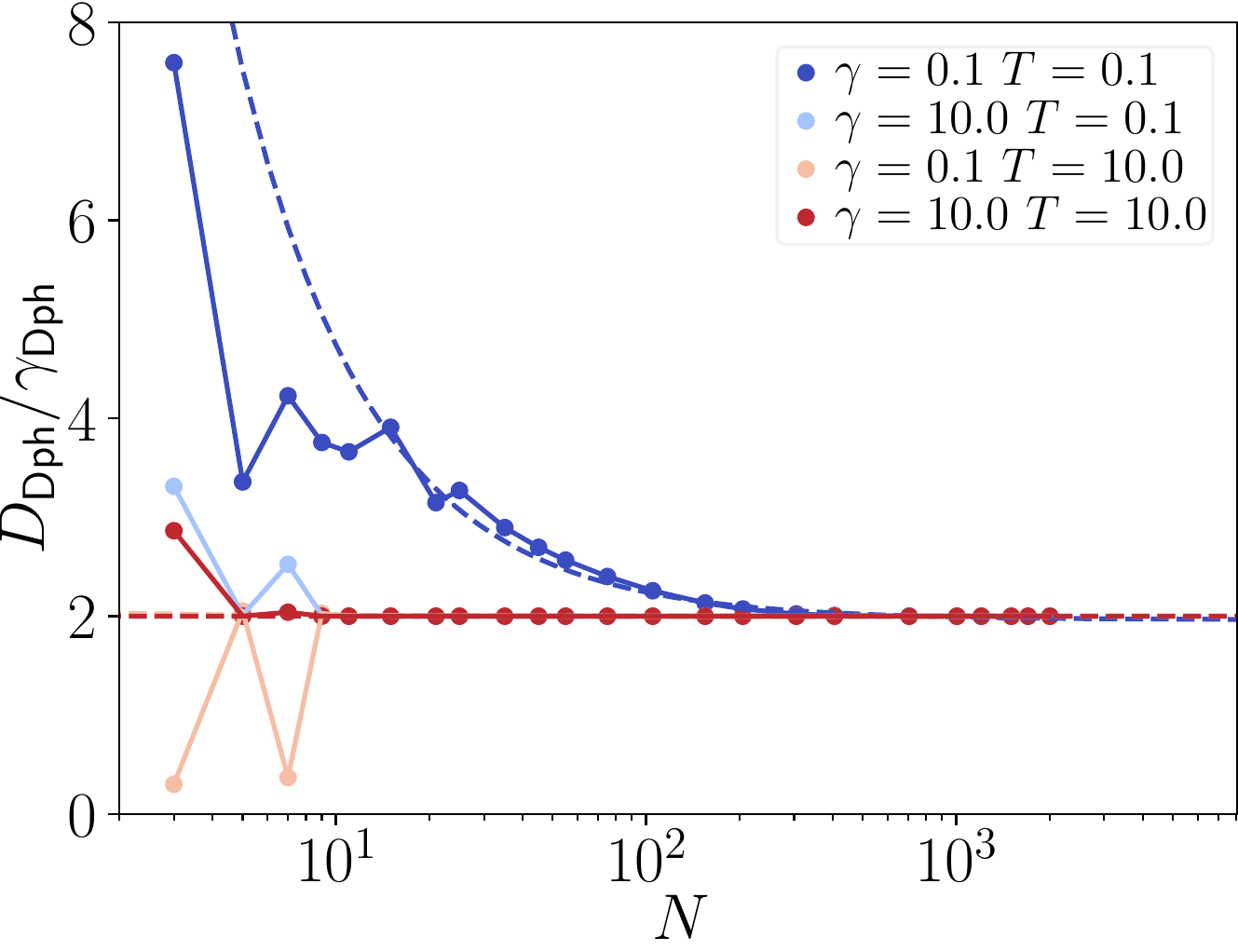}
\par\end{center}
\caption{\label{fig:Dph_finiteN} 
Diffusion constant of the dephasing model at different $(T,\g)$ values. In general, the diffusion constant decays with the inverse system size, which we exploit to extract the $N\ra \infty $ limit from a non-linear fit, dashed lines.}
\end{figure}
Instead, to extract the relevant information in the $N \to \infty$, we use the fact that the diffusion constant has itself a $1/N$ scaling \cite{znidaricNonequilibriumSteadystateKubo2019a} when measured in the middle of the chain. In the QSSEP and dephasing model, we use this result to perform non-linear fits to $D$ as shown in Fig.~\ref{fig:Dph_finiteN}. 

\begin{figure}[h]
\begin{center}
\includegraphics[width=0.95\columnwidth]{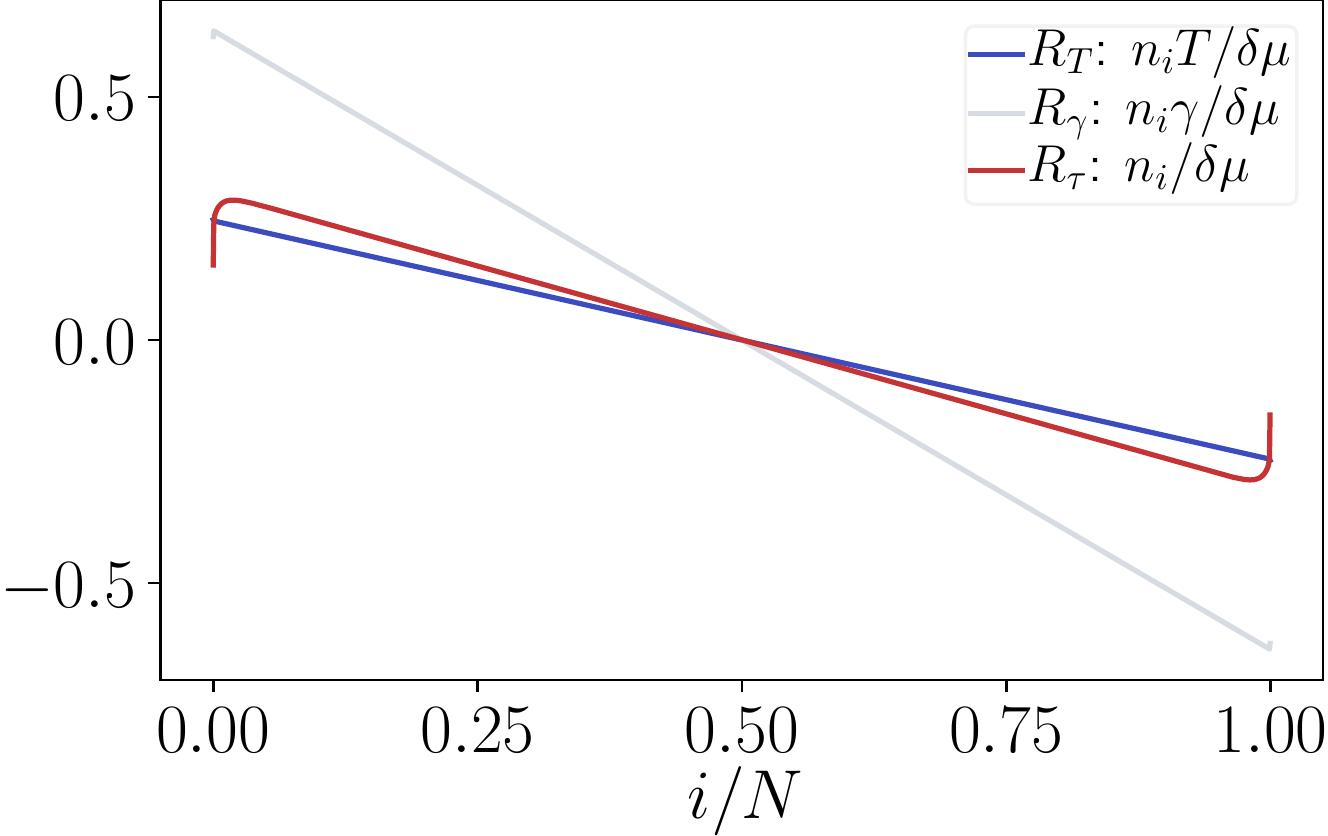}
\par\end{center}
\caption{\label{fig:profile} 
Density profiles of a chain of size $N=2000$ in the diffusive regime for the different regions $R_{\tau,\gamma,T}$ shown in Fig.~\ref{fig:current_phase_diagram-1}. The breaking of Fick's law is limited to a non-extensive number of sites near the edge.}
\end{figure}

In this figure we plot the diffusion constant of the dephasing model as measured in the middle of the chain for increasing system sizes and different $(T,\mu)$ values.  The dashed lines depict the non-linear fit of the function $a+b/(N+c)$ with $a,b,c$ fitting parameters. We find that most observables in these models exhibit $1/N$ corrections as discussed in \cite{znidaricNonequilibriumSteadystateKubo2019a}. The speed of convergence however depends on the point in the phase-space $(T,\mu)$, with region $R_{\tau}$ (see Fig.~\ref{fig:current_phase_diagram-1}) showing the slowest convergence. 
This is a consequence of the effects of the bath discussed in the main text. Deep in the $\tau$-dominated regime, we observe the breaking of Fick's law near the edges as shown in Fig.~\ref{fig:profile}. 

Since this effect only occurs in a finite portion of the system close to the edges, the convergence is only slowed down. We thus evaluate $D$ in the middle of the chain to mitigate its effects and get a better accuracy.

\begin{figure}[h]
\begin{center}
\includegraphics[width=0.95\columnwidth]{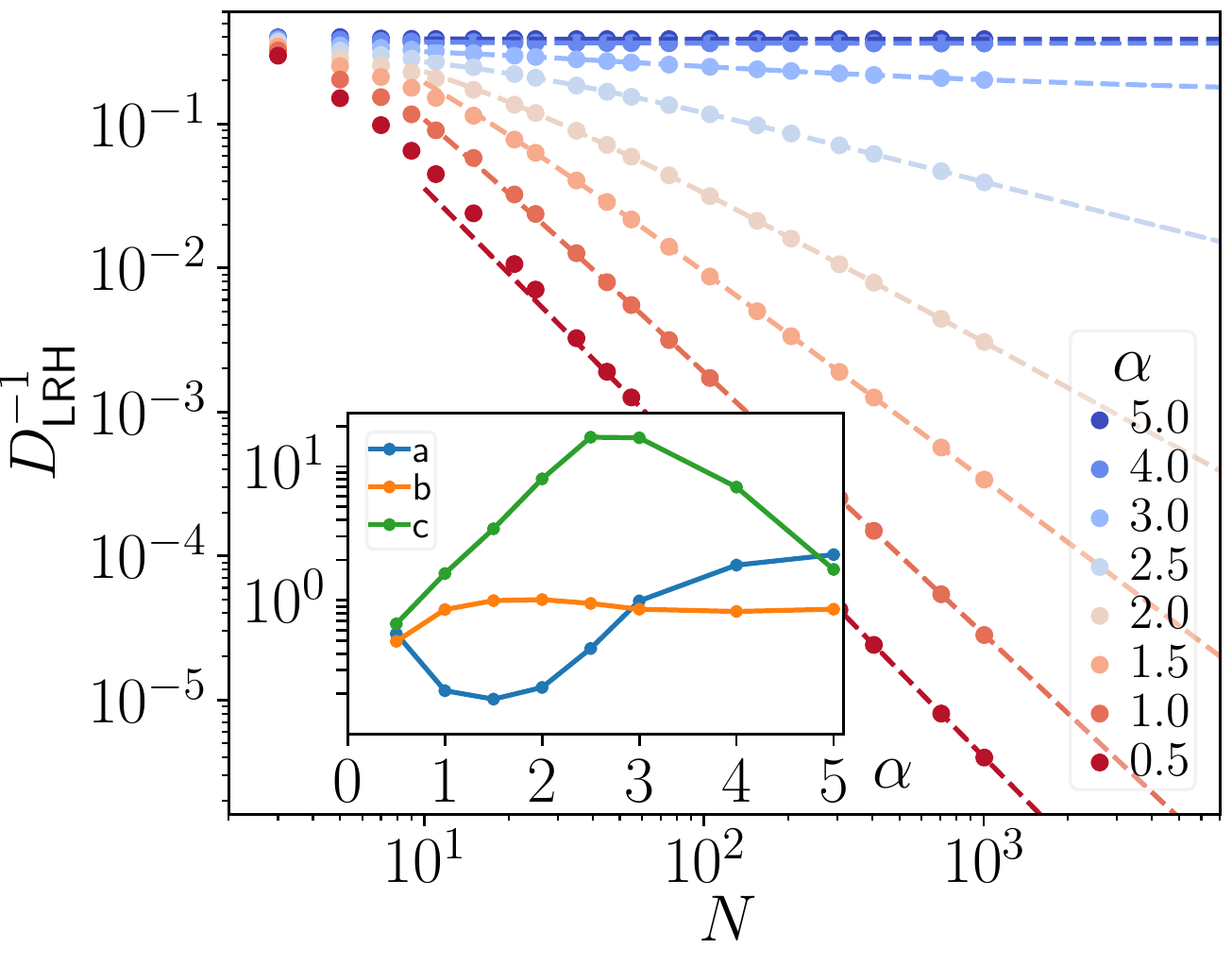}
\end{center}
\caption{\label{fig:DLR_finiteN} 
Inverse of the diffusion constant in the long-range model for different powers of $\alpha$. Dashed lines are fits to $D^{-1}=\left(a H^{(\alpha-b+1)}_{N+|c|}\right)^{-1}$. The results of the fitting are depicted in the inset.}
\end{figure}

For the long-range model, one needs a different approach to obtain the $N \ra \inf$ limit correctly, especially when close to the ballistic-diffusive transition described in the main text. A tentative form for the finite size extrapolation is provided by the solution of the diffusion equation for single particle under a random walk with long-range hopping discussed in Sec.~\ref{sec:LR} which gives a diffusion constant $D=H^{(\alpha-2)}_L$, where $H^{(r)}_x$ is the generalized Harmonic number. 
We find that a fit $D^{-1}=\left(a H^{(\alpha-b+1)}_{N+|c|}\right)^{-1}$, correctly captures the finite-size dependence of $D^{-1}$  for all $\alpha$ values. The fitting parameters $a,b,c$ respectively describe the amplitude, critical exponent and possible finite-size corrections. In Fig.~\ref{fig:DLR_finiteN}, we depict $D^{-1}$ against the result of the fit, respectively dots and dashed lines. 
The best fitting parameters are plotted in the inset. 
The quality of the fit allows us to conjecture that, at the transition point, the diffusion constant diverges logarithmically $D_{LR}(\alpha=\alpha_c)\sim H^{(1)}_N \to \log (N)$.

\section{Coarse-grain length $a$} \label{sec:a-estimate}
In this section, we analytically estimate the coarse-grain length $a$ from the correlation length of the dephasing model. Due to Eq.\eqref{eq:GKinfinity}, it is enough to estimate $a$ from a single Green function, in this case the retarded component.
The starting point is the analytic expression of the elements $G^R_{i,j}$ in the bulk of the chain. For large systems, the boundaries become irrelevant and the good basis of the problem is the momenta basis. In $k$-space, the self-energy takes a diagonal form 
\begin{align}
\S_{k,k'}^{R}	&=\left(-\frac{i\g}{2}\right)\d_{k,k'}.
\end{align}
For the QSSEP, there are cross-diagonal terms in momentum that vanish as $1/L$ and can be safely ignored. Since both self-energy and Hamiltonian are diagonal in the momenta basis, one has
\begin{align}
G^R_{k,k'}&=\d_{k,k'}\frac{1}{\w-\e_k -\S_{k,k'}},
\end{align}
where $\e_k=2\tau \cos(k)$ is the eigenenergy of the bulk Hamiltonian. To find the retarded function in position space, we take the Fourier transform with the continuum limit for $k$ 

\begin{align}
G^R_{r,r'}&=\int \frac{dk}{2\pi} \frac{e^{-ik(r-r')}}{\w - 2\tau \cos k +i\g/2}.
\end{align}

The integral can be solved using the residue theorem and, after some lengthy yet simple manipulations, we find a compact formula 

\begin{align}
G^R_{r,r'}&= \frac{i^{|r-r'|-1}}{2\tau \cos y}e^{iy|r-r'|},
\end{align}
where $y=\arcsin\frac{\w +i\g /2}{2 \t}$ is a complex variable with $\text{Im} (y(\w))>0$. Therefore, in the dephasing model an estimate for the correlation length is given by
\begin{align}
	\xi=\frac{1}{\min \big(\text{Im}\big(\arcsin\frac{\w +i\g /2}{2 \t}\big)\big)} =\frac{1}{\text{arcsinh}\frac{\g}{4 \t}},
\end{align}
In the limit of small dephasing $\g$, we have $\xi=4\tau/\g$ which serves as an estimate for the coarse-grain length $a\sim \tau/\g$. As expected, $a$ should be of the order of the dephasing length $N^* \sim 1/\g$.

\clearpage

\bibliography{main}

\end{document}